# Owning What Can Quit

**Douglas Cumming**
Professor of Finance and Steven Shulman ’62 Endowed Chair for Business Leadership
Stevens Institute of Technology School of Business, 525 River St., Hoboken, NJ 07030, USA
dcumming@stevens.edu

**Acknowledgements and disclaimer:** I received helpful advice from colleagues who preferred to remain anonymous. All errors are mine. The coded dataset of 57 episodes, with sources for every code, and the Stata code for all estimates are available from the author and will be posted with the paper. The account of the Finance Research Letters transition in Section 7 draws on public records and on my own participation as one of the incoming editors; it does not quote correspondence written by others; the only document it describes is the incoming editors’ note to the board, which I co-wrote. The author makes no claims about the intentions, integrity, or conduct of any individual editor, author, journal, or publisher beyond what the cited sources state. Please email me with any corrections or suggestions.

**Funding**
No specific grant from funding agencies in the public, commercial, or not-for-profit sectors.

**Declaration of competing interest**

The author has served in editorial roles for several academic journals, including as co-editor-in-chief of Finance Research Letters (2015–2017), as one of three editors appointed by Elsevier in the 2014 transition described in Section 7; Managing Editor-in-Chief of the Journal of Corporate Finance (2018–2020), succeeding J. Netter; Managing Editor-in-Chief of the British Journal of Management (2020–2022); and Managing Editor-in-Chief of the Journal of Alternative Investments (2025–present), for which he receives a stipend. His term at the Journal of Corporate Finance ended in 2020, when the publisher appointed a new editorial team, and he went on to found the Review of Corporate Finance. He is the founding Managing Editor-in-Chief of the Review of Corporate Finance (2021–present), with NOW Publishers, in connection with which he previously received a 20% ownership interest and royalty rights, but no stipend; NOW Publishers was acquired by Emerald Publishers in May 2025 and the ownership interest was valued at $0. The author has had a dispute with Elsevier arising from a different matter, and has co-authored with one of the other incoming FRL editors named in Section 7.

**Use of generative AI**

In preparing this paper, the author used Claude (Anthropic) for language editing; for drafting and organizing text, including drafts of the model and its proofs; for compiling public sources; for preparing first-pass codes for the 57 Retraction Watch episodes and the editor tenure data from those sources; for writing the Stata code and running the analyses; and for producing the tables and figures. The author set the research question and scope, supplied the account of the FRL transition, directed the analysis, checked the model and its proofs, and reviewed all AI-assisted output. Every code is linked to its public sources in the dataset. The author takes full responsibility for the content, including any errors.

**Data availability**

The coded dataset of the 57 Retraction Watch episodes, the editor tenure dataset, with sources for every code, and Stata do-files that reproduce every table and figure (except the diagram in Figure 1) will be posted with the paper. Journal-level data come from SCImago, as in Cumming (2026a). Other public sources are listed in the references.

# Owning What Can Quit

This draft: November 2026

**Abstract**

Academic journals are part of the infrastructure of science, yet most journal titles are owned by commercial publishers while their standing rests on editors and boards who can leave. This paper asks what that separation means for the governance of science. Using a coded dataset of all 57 editorial mass resignations recorded by Retraction Watch from 2015 to 2026, data on editor tenure at 110 economics, finance and accounting journals, and journal-level publication and revenue data, I document three patterns. First, resignation does not reverse an owner's decision: in none of the 19 episodes involving an editor's replacement did the owner back down after the board resigned, so editorial communities have influence only before they leave. Second, whether a departing board can take a journal's standing with it depends on institutions outside the board: successor journals follow most often where a scholarly society backs the board and rarely in medicine, where non-profit hosts are scarce, while journal ranking lists, which rate titles rather than the people who edit them, keep reputation with the owner. Third, a stated limit on editor tenure was applied at the publisher's discretion; where it was applied, output grew almost fourfold, a short-run gain whose long-run effect on the journal's standing is not yet known. I draw implications for scholarly societies, funders, publishers, and the bodies that compile journal rankings.

**Highlights**

- Codes all 57 editorial mass resignations recorded by Retraction Watch, 2015–2026
- No owner reversed an editor's replacement after the board had resigned
- Successor journals form where societies back boards, rarely in medicine
- Ranking lists rate titles, not editors, so reputation stays with owners
- A stated ten-year cap on editor tenure did not bind at the publisher citing it

## 1. Introduction

In late July 2026, Elsevier declined to renew the contract of the editor-in-chief of *Games and Economic Behavior* (GEB), one of the leading journals in game theory, citing "closer alignment with Elsevier strategic priorities and publishing policies." The remaining editors and 58 of its 62 advisory editors then resigned (Retraction Watch, 2026a). On October 9, 2026, the Game Theory Society, for which GEB had served as an official journal, ended its association with the title and announced that it was exploring a society-owned successor (Game Theory Society, 2026). Elsevier still owns the title, its back catalogue, and its place on ranking lists. The people who made the journal worth reading, and the society that lent it standing, are leaving. Which of the two ends up with the journal's reputation is not yet known.

The episode raises questions about who governs the infrastructure of science. Journals certify findings, allocate attention, and, through the ranking lists used to hire, promote and fund researchers, shape careers and research agendas (Hicks, 2012; Rafols et al., 2012; Mingers and Willmott, 2013; Hicks et al., 2015). Most are owned by a few commercial publishers (Larivière, Haustein, and Mongeon, 2015), while their standing rests on editors and boards whom the publishers do not employ. The questions are also general ones about organization. What does ownership of an asset confer when most of its value lies in people who can quit? When will an owner override those people, and when will it back down? What determines whether they can take the value with them when they leave? And do the rules an owner announces, such as limits on how long a manager may serve, constrain the owner, or give it cover for decisions it would make anyway? In the property-rights theory of the firm, ownership of alienable assets confers residual control rights, the right to decide whatever contracts leave open, while human capital is inalienable and can leave (Grossman and Hart, 1986; Hart, 1988; Hart and Moore, 1990). Power in such organizations rests on access to critical resources (Rajan and Zingales, 1998), and formal authority held by an owner can coexist with real authority held by those with the relevant knowledge (Aghion and Tirole, 1997), and authority that an owner delegates informally lasts only as long as the owner chooses to honour it (Baker, Gibbons, and Murphy, 1999). The theory has shaped how economists think about partnerships, professional-service firms, and the boundaries of the firm, but its central prediction about disputes, that the outcome turns on what the human capital can take with it when it leaves, is hard to test. Control disputes are usually private, and exit is rarely observed in a form that can be measured.

Academic journals offer a revealing setting. The title is an alienable asset, usually owned by a commercial publisher. The journal's reputation, by contrast, is embodied in its editors and board, who are typically not the publisher's employees and who supply nearly all of the scholarly inputs. When the owner and the board disagree, the dispute often becomes public, and when the board leaves, its exit, and any successor journal it founds, are observable. GEB is one such case; Section 5 codes 56 others.

I organize the analysis around a simple model, set out informally in Section 2 and formally in Appendix C. An owner holding the title can replace the editor to pursue a policy that benefits it, such as higher publication volume. The board can protest, and if the owner proceeds, it can stay or leave and launch a successor that takes part of the reputation with it. With complete information, owners withdraw replacements that would trigger costly exit before anyone resigns, so mass resignations occur only when the owner values its policy more than the reputation it loses or, with incomplete information, when it underestimates the board's ability to leave. The model yields five predictions: reversals precede resignations (P1); exit with a successor is more likely where the board can carry the reputation and starting again is cheap (P2); owners enforce general rules selectively, against weak boards (P3); where boards disperse, the owner's policy is implemented (P4); and cheaper hosting of successors raises the frequency of exit (P5).

I test the predictions with three kinds of evidence. The first is a systematic coding of all 57 episodes on the Retraction Watch Mass Resignations List, across every field and publisher from 2015 to 2026 (Retraction Watch, 2026c), which I use to test P1, P2, and P5. The second is a set of episodes in economics and finance, including four that the list omits, which I classify according to the outcomes the model allows. The third is journal-level evidence on P3 and P4: a dataset of chief-editor tenure at 65 Elsevier and 45 other economics, finance, and accounting journals, and publication and citation data before and after two editor replacements. One of those replacements, at *Finance Research Letters* (FRL) in 2014, is an episode in which I was one of three incoming editors, and I disclose my role at the outset of the account in Section 7.

The evidence supports four of the five predictions. On P1, owners almost never reversed course once boards resigned: reversals followed resignations in only three of 57 episodes, all over individual papers, where the owner's gain from holding firm was small, and none involving an editor, while the documented withdrawal of an editor appointment, at the *European Law Journal* in 2018, came before the board resigned (Retraction Watch, 2020). On P2, in a linear probability model, boards of publisher-owned titles affiliated with a society were 43 percentage points more likely to launch or announce a successor than boards of other publisher-owned titles, boards of society-owned titles were 24 points less likely, and boards in medicine and the life sciences, where non-profit hosts are scarce, were 33 points less likely. On P3, a ten-year cap on chief-editor terms that was given as the ground for replacing FRL's founding editor in 2014 did not bind: of 26 Elsevier chief editors with ten or more years in post during 2010–2015, only eight left in that period (31 percent, against 10 percent at other publishers, a difference that is not significant), and a binding cap is rejected even if it was enforced with probability one half ($p = 0.04$). The only departure documented as a non-renewal was at the lowest-revenue Elsevier finance journal in the comparison, which was also unrated in the 2010 ABS guide, and its estimated revenue then grew from $0.37 million in 2013 to $1.86 million in 2018. No editor at a journal in the top grade of the 2010 ABS guide left, against 7 of 16 at lower-graded journals, a difference in the predicted direction that is not statistically significant. On P4, FRL's output rose about 3.8-fold relative to comparable finance journals after the replacement, with no significant change in citations per document. The evidence does not support P5 in its simplest form: the share of episodes producing a successor is the same before and after 2023, although successors are increasingly hosted by non-profit platforms. Editor replacement itself is common and not specific to any publisher: it occurs in a third of the recorded episodes, at each of the four largest commercial publishers and at some societies.

The paper contributes to three literatures. The first is research on the governance of scholarly communication and research evaluation. Studies of research metrics and journal rankings have shown how rankings shape what is published and by whom (Rafols et al., 2012; Mingers and Willmott, 2013; Hicks et al., 2015), how citation indicators used to evaluate researchers and work can mislead (Tol, 2009; Wang, Veugelers, and Stephan, 2017), and how editors, authors and publishers can game them (Wilhite and Fong, 2012; Martin, 2013, 2016; Biagioli, Kenney, Martin, and Walsh, 2019; Alexander and Cumming, 2026). I show that ranking lists also shift power between publishers and editorial communities: because they rate titles rather than the people who edit them, they keep reputation with the owner when a board leaves, and they delay both the entry of successors and the downgrading of titles whose standing falls. The second is the economics of scholarly publishing, which has documented price differences between commercial and non-profit publishers (Bergstrom, 2001), pricing after publisher mergers (McCabe, 2002), the concentration of publication in a few commercial firms (Larivière, Haustein, and Mongeon, 2015), and the editorial process (Ellison, 2002); I add the owner's control over editorial appointments as an instrument of strategy, and companion work on finance journals (Cumming, 2026a, 2026b) to which the FRL episode supplies an earlier stage. The third is the property-rights theory of the firm (Grossman and Hart, 1986; Hart, 1988; Hart and Moore, 1990), whose empirical tests have relied mainly on variation in asset ownership and contractibility (Baker and Hubbard, 2004) and have found it hard to separate property-rights from transaction-cost explanations (Whinston, 2003). Journals separate the alienable asset from the inalienable human capital, make disputes observable, and allow exit to be measured; the evidence shows how complementary institutions, such as scholarly societies and non-profit hosts, govern the outside option of the human capital, and connects to work on spinouts (Franco and Filson, 2006), noncompete agreements (Marx, Strumsky, and Fleming, 2009; Garmaise, 2011), and the allocation of ownership among stakeholders (Hansmann, 1988, 1996; Rajan and Zingales, 2001).

The rest of the paper is organized as follows. Section 2 presents the conceptual framework and its predictions; the formal model and proofs are in Appendix C. Section 3 describes the institutional background, including the case for owner discretion. Section 4 describes the data and the empirical counterparts of the model's parameters. Section 5 tests P1, P2, and P5 on the 57 recorded episodes. Section 6 classifies the economics and finance episodes by the model's outcomes, and Section 7 examines the FRL transition. Section 8 tests P3 with editor tenure data, and Section 9 tests P4. Section 10 summarizes the evidence on the predictions, Section 11 discusses caveats, implications for research policy, and lessons for settings beyond journals, and Section 12 concludes.

## 2. Conceptual framework

The analysis is organized around a simple model of an owner and a board. This section states its elements and results informally; Appendix C gives the formal statements and proofs.

An owner, O, holds the title of a journal whose reputation is worth $R$. The reputation is embodied in the editors and board, B, who act collectively. The owner considers replacing the editor in order to implement a policy, such as higher publication volume or a new review model, that yields the owner a private gain $G > 0$ and imposes a cost $C \geq 0$ on the board, for example by lowering the standing of the work that board members certify. If the board exits, it launches a successor journal at cost $F \geq 0$ and takes with it a share $\theta \in [0, 1]$ of the reputation. The board values reputation under its own control at $w$ per unit, and the owner loses $m$ per unit of reputation that leaves, so the owner's loss from exit is $L = \theta m R$. The cost $F$ captures everything that makes starting again hard: finding a publisher or host, financing, indexing, and coordinating the board. A society that adopts the successor, or a non-profit platform that hosts it, lowers $F$; a society that designates the journal as its own, and moves with the board, raises $\theta$.

The timing is as follows (Figure 1). (1) The owner announces a replacement. (2) The board may protest. (3) The owner proceeds or withdraws. (4) If the owner proceeds, the board stays or exits; exit is irreversible once the successor is launched. With the status quo normalized to zero, the payoffs to the owner and the board are (0, 0) if the owner withdraws; $(G, -C)$ if it proceeds and the board stays; and $(G - L, \theta w R - F)$ if it proceeds and the board exits.

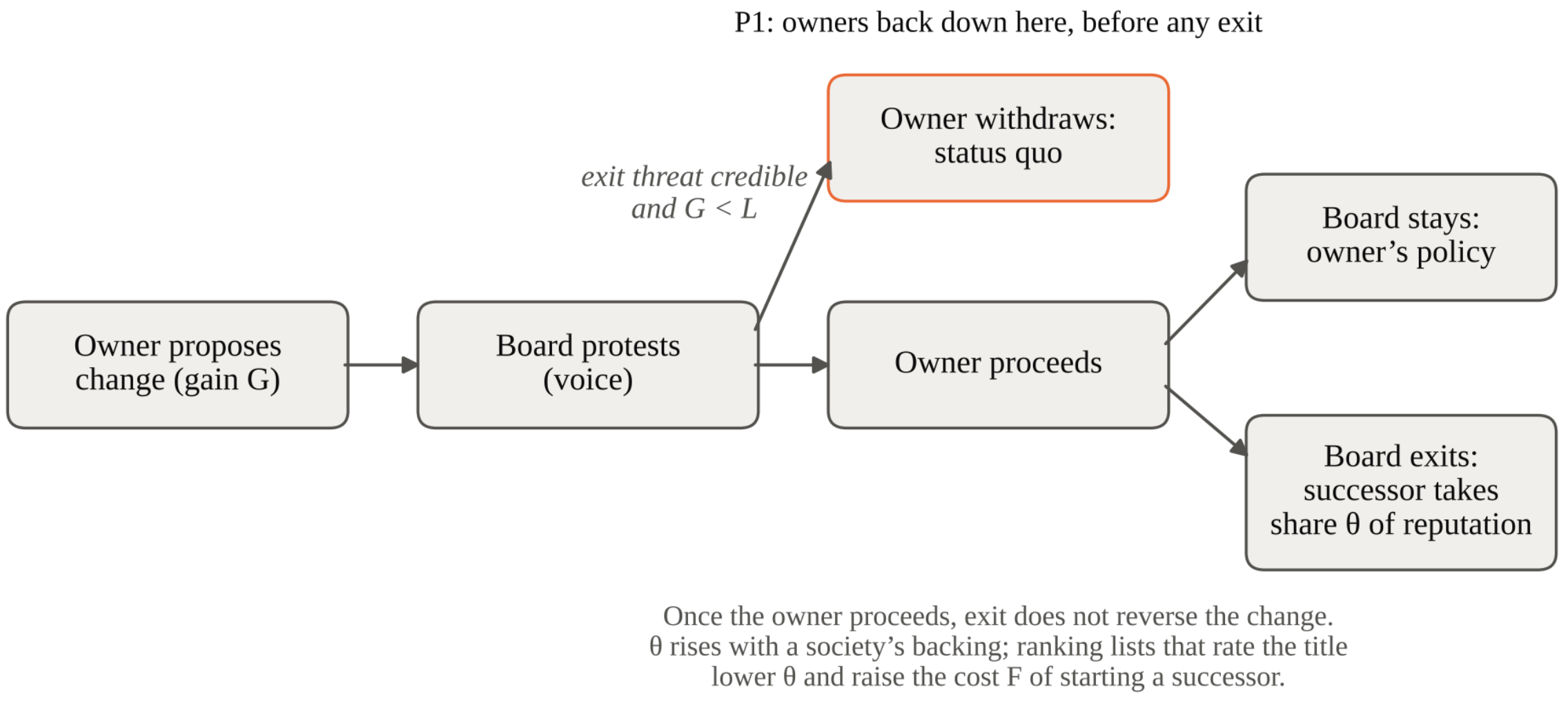

**Figure 1: Timing of the game and where the predictions apply**

Note: The owner, who holds the title, proposes a change of editor or policy with gain G. The board can protest. If the board's threat to exit is credible and the owner's gain is smaller than the reputation it would lose (G < L), the owner withdraws; otherwise it proceeds, and the board then stays or exits and founds a successor that takes a share θ of the reputation. P1 concerns the withdrawal branch; P2 and Section 3.5 concern the determinants of θ and of the cost F of starting a successor.

Six results follow. First, with complete information, an owner withdraws a contested change before anyone resigns whenever the board's threat to leave is credible and the owner's gain is smaller than the reputation it would lose; mass resignations occur only when the owner values its policy more than that reputation (Proposition 1). Second, when the owner cannot observe the board's cost of starting again, it may misjudge the board and proceed; among the changes that go ahead, the probability that the board leaves rises with the share of the reputation it can take, and comparisons across observed episodes recover this effect even though the owner chooses which changes to make (Proposition 2 and Lemma 1). Third, a general rule, such as a limit on editor tenure, that the owner enforces at its discretion will be applied where boards are weak or the owner's gain is high, and waived elsewhere (Corollary and Proposition 3). Fourth, because a successor journal needs enough of the board to leave together, exit is a coordination problem; a society or host that commits in advance to support a successor removes the risk of a failed exit, so departures should be either near-total with a successor or scattered without one (Proposition 4). Fifth, public, collective protest works when it commits the board to leave, which widens the set of boards whose objections succeed (Proposition 5). Sixth, if there is a window between resignations and the launch of a successor, an owner may reverse course after resignations, but only where its gain from the change is small relative to the reputation it can recover (Proposition 6).

### *2.1 Predictions*

The model yields five predictions. **P1 (timing of reversals).** Owners withdraw contested changes before boards resign, typically after public, collective protest. Reversals after resignation are rare, confined to disputes in which the owner's gain is small, such as disputes over individual papers, and absent where the change is a new editor chosen to carry out a policy (Propositions 1, 5, and 6). **P2 (outside options).** Collective exit with a successor is more likely where the board can take a large share of the reputation, as when a society that treats the journal as its own moves with the board, or where starting again is cheap, as where non-profit hosts exist (Proposition 2(b) and (c)); by Lemma 1, these comparisons can be made among the episodes in which the owner proceeded. Where an organization commits in advance to host a successor, exits should be collective rather than piecemeal (Proposition 4). Where the society itself owns the title, its brand stays with the title, $\theta$ is low, and successors should be rare. **P3 (rules on tenure).** A cap on tenure enforced at the owner's discretion will not bind: most long-serving editors remain, and those replaced are at journals with weak boards (Corollary; Proposition 3). **P4 (policy after replacement).** Where a replacement proceeds and the board stays or disperses, the owner's policy is implemented, so outcomes such as publication volume should move toward the owner's objective (Proposition 1(a) and the definition of $G$). **P5 (waves of exit).** A fall in the cost of launching successors, such as the growth of non-profit and diamond open-access hosts, raises the probability of exit at journals where the owner proceeds (Proposition 2(c)).

Because it also deters the replacements most likely to provoke exit, the effect on the share of observed episodes ending in exit is ambiguous.

## 3. Institutional background

### *3.1 Who owns the title*

Economics and finance journals fall into three ownership types. A society can own the title and contract with a publisher for production and distribution; the society then appoints the editor, and if it is unhappy with the publisher it can move the title. A publisher can own the title outright and contract with editors for fixed, renewable terms. And a hybrid arrangement is common, in which a society designates a publisher-owned journal as its official journal, or founded the journal but does not own it. In the hybrid case the society's influence over appointments rests on convention rather than contract: it is informal authority in the sense of Baker, Gibbons, and Murphy (1999), which lasts only as long as the owner finds it worth honouring.

The episodes below turn on this distinction. The European Economic Association had adopted the *European Economic Review* (EER) as its official journal, but the title belonged to Elsevier. When the association broke with Elsevier over pricing, it could not take the EER with it; it replaced the EER as its official journal with a new title, the *Journal of the European Economic Association* (JEEA), launched with MIT Press and first published in 2003, while Elsevier continued to publish the EER with no tie to the association (Chronicle of Higher Education, 2003). The *Journal of Asian Economics* (JAE) was founded in 1990 by the American Committee on Asian Economic Studies (ACAES), whose leadership selected each editor-in-chief "with the concurrence of the publisher/owner of the Journal" (ACAES, n.d.). GEB is one of the Game Theory Society's two official journals, but is owned by Elsevier. In each hybrid case the society's role in choosing the editor could be ended by the owner.

### *3.2 Editor contracts and the stated ten-year cap*

Editors of publisher-owned journals serve under fixed-term contracts that the publisher may decline to renew. I am not aware of a published Elsevier policy limiting the tenure of chief editors. As an incoming editor of FRL in 2014, however, I was told that Elsevier's policy at the time limited chief editors to ten years, on the grounds that new editors bring new ideas and long tenures tend to settle into the status quo. The cap applied to chief editors in post around 2010–2014; new appointments, including my own in 2014, were made on three-year renewable terms. In the GEB episode, an Elsevier spokesperson said that Moulin's contract had "reached the end of its three-year timespan" (Retraction Watch, 2026a). Short renewable terms give the owner frequent occasions to decide whether to renew, which in the framework of Section 2 are the occasions on which a replacement can be proposed. Section 8 compares the stated cap with the tenure of Elsevier's chief editors.

### *3.3 What boards can do*

An editorial board facing a succession it opposes has three responses. Its members can resign, individually or together. They can threaten to resign, which is an exercise of voice in Hirschman's (1970) sense. And they can leave and start a successor journal, which is exit in its strongest form. Hirschman's central point, that voice is effective when exit is credible, is formalized in Section 2 and Appendix C: an owner takes a threat seriously if the board could take the journal's reputation elsewhere, and not if its members can only leave one by one.

### *3.4 The owner's case*

There is a serious case for owner discretion, and the analysis below does not depend on rejecting it. The owner carries the financial risk, the legal liability, and the long-run reputation of the title, while boards are rarely accountable for a journal's finances. Fixed, renewable terms for editors are common at society journals as well as commercial ones, and term limits are a standard governance device for renewing leadership and avoiding entrenchment. A board may be largely inactive. And a journal that grows after a change of editor may be meeting real author demand. The question here is therefore not whether owners may replace editors, which they plainly may, but how the right is used, on what stated grounds, and what happens when the board objects.

### *3.5 What the title is worth, and why reputation sticks to it*

The owner's stake in the title is large. Elsevier's parent, RELX, reports that its Scientific, Technical and Medical segment earned an adjusted operating profit of £1,035 million on revenue of £2,714 million in 2025, excluding print, which RELX now reports separately, a margin of 38 percent (RELX, 2026). At the level of single journals, Cumming (2026a) estimates revenue for 26 finance journals. On those estimates, FRL earned about $0.37 million in 2013, the year before its founding editor's contract ended, the least among the Elsevier finance journals in Table 6 (the *Journal of Economics and Business* $0.46 million, the *Journal of Financial Stability* $0.54 million, the *Pacific-Basin Finance Journal* $0.64 million, the *International Review of Economics & Finance* $1.14 million, the *Journal of Corporate Finance* $1.29 million, and the *Journal of Financial Economics* $3.20 million).

In the model these magnitudes enter twice: the owner's loss if the board leaves, $L = \theta mR$, rises with the journal's revenue $R$, and the owner's gain from a change of policy, $G$, is largest where a title has room to grow.

Two institutions tie reputation to the title rather than to the board, and so lower $\theta$ or raise the cost $F$ of starting a successor. The first is contractual. When the European Economic Association left the EER, two former EER editors could move to the new journal only once the noncompete clauses in their Elsevier contracts expired (Chronicle of Higher Education, 2003). A noncompete clause raises $F$ directly, by delaying the people who carry the reputation; Garmaise (2011) shows that stronger enforcement of such agreements reduces executive mobility and changes both executive compensation and firm investment.

The second, and more pervasive, is the journal ranking list. The lists that business schools and economics departments use for hiring, promotion, and pay, such as the Academic Journal Guide of the Chartered Association of Business Schools (formerly the ABS guide), the Australian Business Deans Council list, and the *Financial Times* 50, rate titles, not editors or boards. When a board leaves, the rating stays with the title, and a successor must earn its own, which takes years because lists are revised only every few years and usually require a citation record. Authors whose careers depend on the lists have little reason to follow a board to an unrated successor, and board members, who are themselves authors and evaluators, face the same incentive. Ranking lists therefore make part of a journal's reputation alienable: they transfer it from the people who built it to the title the owner holds. In the model they lower $\theta$ and raise $F$, which strengthens the owner's residual control and makes boards less willing to leave, even when they object to a change.

The 2010 ABS guide illustrates both the lag and the way a society can shorten it. It did not rate FRL or the *Journal of Financial Stability*, both founded in 2004, or the *International Review of Economics & Finance*, founded in 1992 (Association of Business Schools, 2010). It did rate the JEEA, launched in 2003 with the backing of the European Economic Association, at grade 4 after only seven years, up from grade 3 in the 2009 edition and one grade above the EER, which the association had left. A society's backing appears to let a successor clear the ranking hurdle quickly, a further channel through which societies raise $\theta$ (P2). The EER case also shows the cost of an exit to the owner: Elsevier kept the title, but within seven years the journal it had lost the society to was ranked above it.

## 4. Data and empirical counterparts of the model

### *4.1 The Retraction Watch episodes*

The Retraction Watch Mass Resignations List records episodes in which editors of a journal, or of a book programme, resigned together in a dispute (Retraction Watch, 2026c). As of October 2026 it lists 57 episodes from 2015 to 2026. I code every episode on the list. For each, I record the field; who owned the title at the time of the dispute (the publisher, a society, or an independent body); whether the journal was owned by, or was the official journal of, a scholarly society; the primary and secondary triggers; whether the publisher or owner removed, declined to renew, or replaced the editors at any point in the episode; whether the departing editors or the society launched or planned a successor journal, and if so its host; whether the owner reversed or materially changed the disputed decision; and whether the original title continued. Each code is supported by one or more public sources, listed in the dataset. Appendix A reports the codes.

Three coding conventions matter for the results. Editor replacement is coded only when the owner pushed editors out, including by declining to renew them or by appointing successors over the board's objections; appointments made after editors had already resigned are not counted, since otherwise nearly every episode would qualify. Successor journals are coded as launched once announced with a named title and host, and as planned when a society or the departing editors have announced an intention without a title. Ownership is often not stated in public sources; where an owner acted on its own authority and no other owner appears, I code the publisher as owner and flag the inference in the dataset.

### *4.2 Economics and finance episodes*

The list starts in 2015, depends on press coverage, and includes four economics journals and no finance journals. I add four episodes from economics and finance that meet one rule: an owner-initiated change in the editorship, or in a society's role in choosing the editor, that was contested publicly by the editor, the board, the society, or other stakeholders. They are documented in news reports, society and university records, and, for FRL, my own participation: the *European Economic Review* (2003), the *Journal of Banking & Finance* (2008), FRL (2014), and the *Journal of Asian Economics* (2020). Because these were found partly through my own knowledge of finance journals, most of which are published by Elsevier, they are used to illustrate the model's outcomes, and not to compare publishers.[1]

### *4.3 Empirical counterparts of the model*

Table 1 maps the model's parameters and outcomes to observable features of the episodes. The share of reputation the board can take, θ, is high when a scholarly society treats a publisher-owned journal as its own and can move with the board, and low when the society itself owns the title, since the society's brand then stays with the title. The cost of starting again, *F*, is lower where non-profit, diamond open-access, and university-press hosts are established, as in the humanities, social sciences, and mathematics, and higher where they are scarce, as in medicine and the life sciences. The owner's gain, *G*, is reflected in triggers concerning publication volume and fees. Each of the model's outcomes has an observable counterpart: a change of editor or policy that goes ahead without a successor, even if board members resign (case a), a withdrawal before resignation (case b), and exit with a successor (case c).

**Table 1: Empirical counterparts of the model**

| Model object | Interpretation | Empirical counterpart |
|---|---|---|
| θ (share of reputation the board can take) | Whether the reputation can move with the board | High: publisher-owned title affiliated with a society. Low: society-owned title |
| *F* (cost of starting again) | Hosting, financing, indexing, coordination | Lower in fields with established non-profit hosts; higher in medicine and life sciences |
| *G* (owner's gain from its policy) | Revenue from volume or fees | Triggers: volume pressure, fees and open-access policy; FRL operating plan |
| *C* (board's cost of staying) | Scholarly cost of the owner's policy | Triggers: editor replacement, editorial interference |
| Case (a): proceed, board stays or disperses | Replacement or policy goes ahead | Owner's change of editor or policy proceeds; board members may resign but no successor follows |
| Case (b): owner withdraws | Voice succeeds | Reversal before resignation |
| Case (c): proceed, board exits | Exit | Successor launched or planned |

[1] I do not include the change of editors at the *Journal of Corporate Finance* in 2020, where I was one of the editors, because it does not meet the selection rules of any sample in the paper. It is not on the Retraction Watch list, and the change, which took effect when the editors' terms ended in 2020, was not contested publicly by the editors, the board, or anyone else, so it does not meet the rule for the episodes added in this section; and my term (2018–2020) began after 2013 and lasted less than ten years, so it is outside the tenure sample in Table 6, where the journal's two earlier long-serving editors appear. The *Review of Corporate Finance*, which I founded in 2021 with NOW Publishers, is a new journal rather than a successor launched by a departing board. Including the episode anyway, coded as an editor change at a publisher-owned title without a society, moves the coefficient on society affiliation in Table 4, column (3), from 0.429 to 0.448 if the *Review of Corporate Finance* is not counted as a successor and to 0.410 if it is; no other coefficient moves by more than 0.04, and the Fisher test for society affiliation in Table 3 stays at p = 0.10. Treating 2021, the first year under the new editors, as an event year in the design of Section 9, with a window that includes the ecosystem period and so is only descriptive, the journal's output did not change relative to the control journals (−0.11 log points, placebo p = 0.67), while its citations per document rose (2.22, placebo p = 0.07). My roles at both journals are disclosed in the declaration of competing interest.

Note: Model objects from Section 2. Codes for each episode are in Appendix A.

### *4.4 Editor tenure data*

To test P3, I compile the tenure of the editors in the top editorial position, the editor-in-chief, managing editor, or co-equal lead editors, at 65 Elsevier economics, finance, and accounting journals and 45 comparison journals published by Wiley, Oxford University Press, Springer, Cambridge University Press, Taylor & Francis, university presses, and societies. For each journal I record every spell in the top position active at any time from 2005 to 2020, with start and end years, the end reason where stated, the owner, and society affiliation, from editors' curricula vitae, society and university announcements, journal editorials, and RePEc and Wikipedia listings. Across two rounds of collection, the dataset records 549 spells; for 26 Elsevier editors and 10 others who meet the test criteria in Section 8, tenure through 2015 is verified. Publisher websites were not accessible for verification, so coverage is best for journals whose editorial histories are documented elsewhere. Appendix B describes the dataset.

## 5. Evidence from 57 recorded episodes: P1, P2, and P5

### *5.1 What owners propose*

In the model, an owner proposes a change when its gain *G* is large relative to the expected loss of reputation. Table 2 summarizes the episodes by publisher. The most common primary triggers are open-access policy and fees (14 episodes), editor replacement (13), editorial interference by the publisher or owner (10), and pressure to raise publication volume (7). Fees and volume are the triggers associated with the owner's gain *G*; editor replacement and editorial interference mark a cost *C* to the board (Table 1). Counting secondary triggers, the owner removed, did not renew, or replaced editors in 19 episodes (33 percent). Editor replacement is not confined to one publisher: it appears in four episodes each at Elsevier, Wiley, and Taylor & Francis, in two at Springer Nature, and in five at other publishers and owners; in three of the 19, the owner was a society or professional association acting as owner of its own journal. Because the list depends on press coverage and the publishers differ in size, these counts should not be read as rates.

**Table 2: Retraction Watch mass resignations by publisher, 2015–2026**

| Publisher | Episodes | Editor replacement involved | Successor launched or planned | Publisher reversal |
|---|---|---|---|---|
| Elsevier | 8 | 4 | 6 | 0 |

| **Publisher** | **Episodes** | **Editor replacement involved** | **Successor launched or planned** | **Publisher reversal** |
|---|---|---|---|---|
| Wiley | 16 | 4 | 7 | 1 |
| Springer Nature | 9 | 2 | 5 | 1 |
| Taylor & Francis | 7 | 4 | 3 | 1 |
| MDPI | 3 | 0 | 0 | 0 |
| Other publishers and owners | 14 | 5 | 4 | 0 |
| **All** | **57** | **19** | **25** | **3** |

Note: Episodes on the Retraction Watch Mass Resignations List as of October 2026. Editor replacement = the publisher or owner removed, declined to renew, or replaced the editors at any point in the episode. Publisher reversal = the owner reversed or materially changed the disputed decision after the protest. Codes and sources are in Appendix A and the dataset.

### *5.2 P1: reversals precede resignations*

Proposition 1 implies that, with complete information, an owner never reverses a replacement after the board has resigned, and Proposition 6 that, with incomplete information, reversals after resignation occur only where the owner's gain from holding firm is small. The evidence matches. Owners reversed course after a mass resignation in only three episodes, and none of the three concerned an editor: at *Scientific Reports* a disputed paper was later retracted; at *Third World Quarterly* a disputed essay was withdrawn, though the stated reason was threats to the editor; and at *Molecular Genetics & Genomic Medicine* the disputed papers were retracted more than two years after the resignations. In none of the 19 editor-replacement episodes did the owner reinstate the editor or withdraw the replacement after the board resigned.

Proposition 6 explains the pattern. All three reversals after resignation concerned individual papers, where the owner's gain from holding firm was small relative to the reputation at stake, and none of the 19 editor replacements, where the owner's gain was a policy carried out by a new editor, was reversed. The difference (0 of 19 against 3 of 38) is in the predicted direction but too small to be significant (Fisher exact $p = 0.54$).

Case (b) of Proposition 1, in which voice succeeds, predicts that withdrawals occur before resignations, so they should be largely absent from a list of resignations. The documented case fits. At the *European Law Journal*, Wiley sought in 2018 to appoint editors-in-chief without consulting the board of editors; after protest, it agreed to an open competitive process run by a committee of board representatives (Retraction Watch, 2020). The board resigned only later, in 2020, over the terms of a new contract. The successful protest was public and collective and came before any resignation, as Proposition 5 predicts; whether members pledged to resign is not documented. A list of mass resignations is therefore, in the model's terms, a list of cases (a) and (c): replacements that went ahead.

### *5.3 P2: outside options and exit*

Proposition 2(b) and (c) predict that exit with a successor is more likely where the board can take a larger share of the reputation, $\theta$, and where starting again is cheap, a low $F$. Table 3 and Figure 2 report successor rates by the empirical counterparts of $\theta$ and $F$. Where a society was affiliated with a title owned by the publisher, a high-$\theta$ case, a successor followed in six of seven episodes (86 percent), typically with the society's backing: *Indoor Air* (succeeded by the society-owned *Indoor Environments*), *Intelligence* (*Intelligence & Cognitive Abilities*, now the official journal of the International Society for Intelligence Research), *Mathematical Logic Quarterly* (*ZML*, published by the German logic society), *Kurdish Studies* (*Kurdish Studies Journal*), and GEB and *Design Studies*, where the societies have announced plans. Where no society was involved, a successor followed in 17 of 38 episodes (45 percent). Where the society itself owned the title, a low-$\theta$ case, a successor followed in one of eight. By field, successors followed in 3 of 19 episodes in medicine and the life sciences, a high-$F$ setting, against 22 of 38 elsewhere.

**Table 3: Successor journals by ownership, field, and trigger**

| | Episodes | Successor launched or planned | Share |
|---|---|---|---|
| **Panel A: Ownership of the title (θ)** | | | |
| Publisher-owned, society-affiliated (high θ) | 7 | 6 | 86% |
| Publisher-owned, no society | 38 | 17 | 45% |
| Society-owned (low θ) | 8 | 1 | 13% |
| Independent or unclear | 4 | 1 | 25% |
| **Panel B: Field (F)** | | | |
| Humanities and social sciences | 19 | 11 | 58% |
| Physical sciences, engineering, and multidisciplinary | 8 | 5 | 63% |
| Mathematics and statistics | 7 | 4 | 57% |
| Economics and finance | 4 | 2 | 50% |
| Medicine and life sciences (high F) | 19 | 3 | 16% |
| **Panel C: Trigger** | | | |
| Editor replacement involved | 19 | 6 | 32% |
| Other triggers | 38 | 19 | 50% |
| **All episodes** | **57** | **25** | **44%** |

Note: Fisher exact tests: publisher-owned society-affiliated versus publisher-owned without a society, p = 0.10; society-owned versus all others, p = 0.07; medicine and life sciences versus all other fields, p = 0.004; editor replacement versus other triggers, p = 0.26. Source: author's coding of the Retraction Watch Mass Resignations List; see Appendix A.

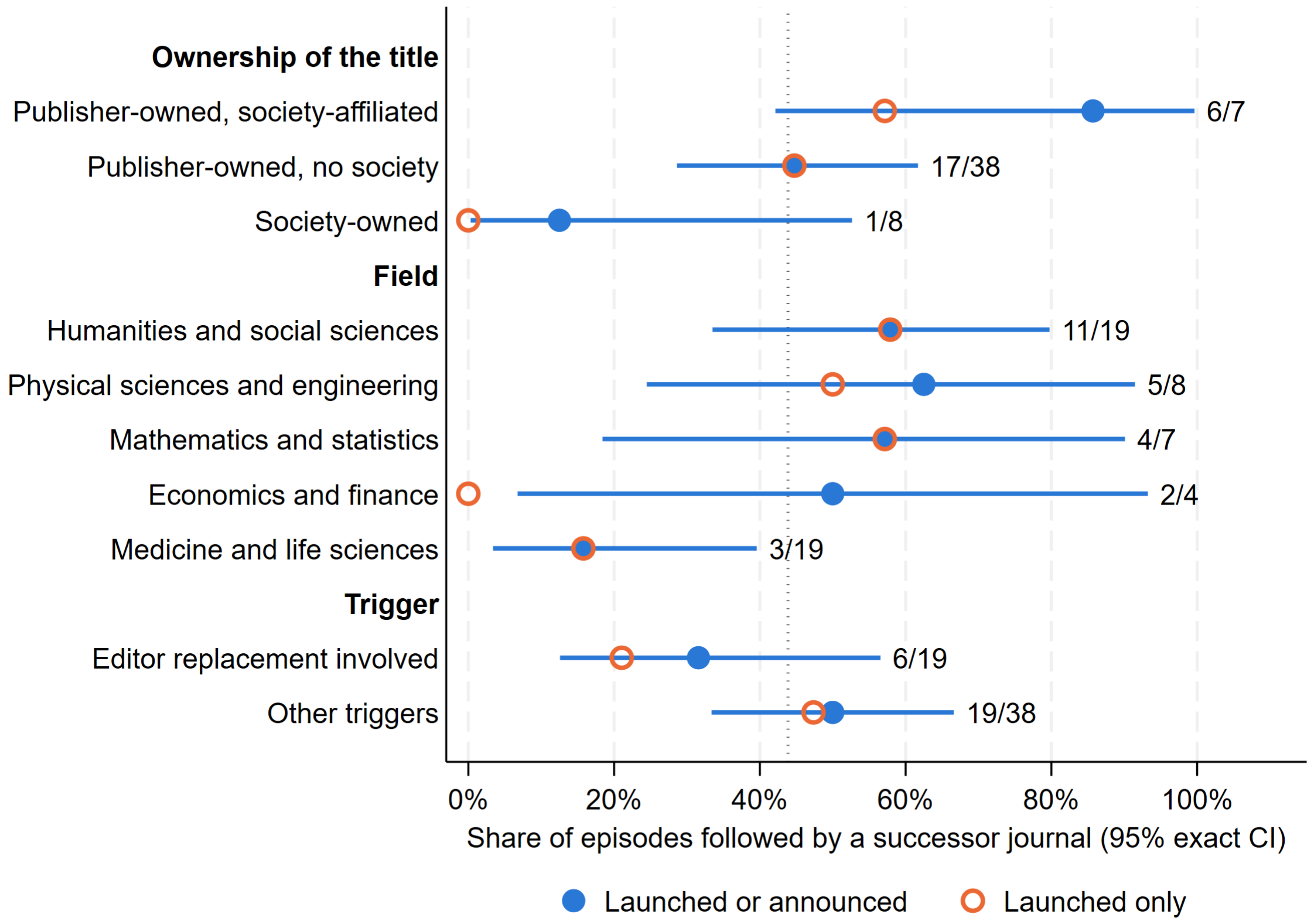

**Figure 2: Successor journals by ownership, field, and trigger, 57 episodes**

Note: Filled dots are the share of episodes in which a successor journal was launched or announced, with 95 percent exact (Clopper–Pearson) confidence intervals; hollow dots are the share in which a successor had been launched by October 2026. Counts at right are successors launched or announced over episodes. The dotted line is the share for all 57 episodes. Source: Appendix A; Table 3.

Table 4 estimates the same relationships jointly in a linear probability model of whether a successor was launched or planned. Relative to publisher-owned titles without a society, society affiliation raises the probability of a successor by 43 percentage points (robust standard error 0.13), and society ownership lowers it by 24 points (0.14). Episodes in medicine and the life sciences are 33 points less likely to produce a successor (0.14). Conditional on these, episodes involving editor replacement are no more likely to produce a successor, and the period after 2023 shows no difference. All three coefficients on $\theta$ and $F$ have the signs the model predicts. Editor replacement, which Table 1 maps to a high cost of staying $C$ and so to more exit, has the opposite sign, although the estimate is not significant. With 57 episodes and only seven in the society-affiliated group, the estimates should be read as descriptive, and the robust standard error for that group, in which six of seven outcomes equal one, is unreliable; the exact test in Table 3 ($p = 0.10$) is the better guide.

**Table 4: Linear probability models of successor launch**

| | (1) | (2) | (3) |
|---|---|---|---|
| Publisher-owned, society-affiliated (high θ) | 0.429∗∗∗ (0.157) | 0.360∗∗∗ (0.126) | 0.429∗∗∗ (0.125) |
| Society-owned (low θ) | −0.304∗∗ (0.143) | −0.244∗ (0.141) | −0.238∗ (0.139) |
| Medicine and life sciences (high F) | | −0.357∗∗∗ (0.127) | −0.334∗∗ (0.143) |
| Editor replacement involved | | | −0.156 (0.123) |
| Episode in 2023 or later | | | −0.101 (0.126) |
| Constant | 0.429∗∗∗ (0.078) | 0.548∗∗∗ (0.092) | 0.652∗∗∗ (0.122) |
| Observations | 57 | 57 | 57 |
| $R^2$ | 0.14 | 0.26 | 0.29 |

Note: Dependent variable equals one if the departing editors or the society launched or planned a successor journal. The omitted ownership category is publisher-owned titles without a society; independent and unclear owners are included in the omitted category. Heteroskedasticity-robust (HC1) standard errors in parentheses. ∗∗∗ $p < 0.01$, ∗∗ $p < 0.05$, ∗ $p < 0.10$. Columns: (1) ownership only; (2) adds field; (3) adds trigger and period. Source: Appendix A.

Two robustness checks qualify the society-affiliation result. First, two of the six successors in that group, at GEB and *Design Studies*, had been announced but not launched by October 2026. Counting only launched successors, the society-affiliated share falls to four of seven, against 17 of 38 for titles without a society (Fisher $p = 0.69$), and the coefficient in column (3) of Table 4 falls to 0.18 (0.17); the coefficients on society ownership (−0.37, standard error 0.09) and on medicine and the life sciences (−0.26, 0.14) remain negative and significant at 10 percent or better, and the Fisher test for medicine gives $p = 0.02$. The society-affiliation result therefore depends on whether announced successors materialize. Second, restricting to the 35 episodes whose ownership is documented rather than inferred, the ownership pattern has the same direction (successors in three of four society-affiliated publisher-owned titles, 11 of 21 without a society, and one of seven society-owned titles, which together account for 32 of the 35), but the samples are too small to distinguish the differences from chance. The hosts of successors point to the same mechanism: of the 23 successors whose host is known, 12 are published on non-profit or diamond open-access platforms such as the Open Library of Humanities and Centre Mersenne, five by university presses, one by a society, and five by commercial publishers.

### *5.4 P5: cheaper hosting and the frequency of exit*

Proposition 2(c) predicts that a fall in the cost of starting again raises the frequency of exit among replacements that proceed. As Section 2.1 notes, the same shift also deters the replacements most likely to provoke exit, so the share of observed episodes ending in exit need not rise. The evidence does not show a rise. Successors followed in 8 of 18 episodes before 2023 and in 17 of 39 from 2023 onward, 44 percent in both periods, and the coefficient on the later period in Table 4 is small and insignificant. What has changed is where successors go: non-commercial hosts account for five of eight successors with known hosts before 2023 and 13 of 15 since. The number of episodes on the list rose sharply after 2023, but because the list itself was started in 2023 and depends on coverage, that increase cannot be attributed to lower costs of exit.

## 6. Economics and finance episodes and the model's outcomes

Table 5 classifies the four economics episodes on the Retraction Watch list and the four that it omits by the model's outcomes. Cases (a) and (c) both appear; case (b), a withdrawal before anyone resigns, rarely reaches a list of resignations. The European Economic Review is case (c) in its clearest form: a society that could carry the reputation (high θ) left with the board's support and founded a successor. The Journal of Banking & Finance, FRL, and the Journal of Asian Economics are case (a): replacements that went ahead with no society able to host a successor, and in the JAE's case a society that existed but did not own the title. GEB is moving toward case (c), with a society that has ended its association and is exploring a successor. The three non-Elsevier episodes on the list concern strategy, publication volume, and contract terms rather than the removal of an editor; at the *Review of Social Economy*, where the association itself owns the title, the departing editors plan a new journal with a university press, the one successor among the eight society-owned titles in Section 5. The four replacements of a chief editor in Table 5 are all at Elsevier; as noted in Section 4, this partly reflects how the off-list episodes were found, and Section 5 shows that editor replacement occurs at each of the four largest commercial publishers.

**Table 5: Economics and finance episodes and the model's outcomes, 2003–2026**

| Journal | Publisher | Year | Trigger | Board response and outcome | Model case | Source |
|---|---|---|---|---|---|---|
| European Economic Review | Elsevier | 2003 | Society dispute with publisher over pricing | JEEA (MIT Press) replaces EER as the society's official journal; EER continues at Elsevier | (c), high θ | Public |
| Journal of Banking & Finance | Elsevier | 2008 | Co-founding editor of about 30 years succeeded; | New editor installed; no successor | (a) | Public |

| Journal | Publisher | Year | Trigger | Board response and outcome | Model case | Source |
|---|---|---|---|---|---|---|
| | | | office moved from Europe to US | | | |
| Finance Research Letters | Elsevier | 2014 | Founding editor not renewed; ten-year cap cited | Board resignations; new editors installed; volume expanded; no successor | (a) | Public; author |
| Journal of Asian Economics | Elsevier | 2020 | Editor-in-chief given one-year terminal contract; society's appointment role ended | Stakeholder letters; publisher kept decision | (a), society without title | Public |
| Econometrics | MDPI | 2023 | Editors and publisher could not agree on strategy | Editors resign; journal continues; no successor | (a), policy change | Public |
| Journal of Economic Surveys | Wiley | 2024 | Concern that publisher emphasized quantity over quality; new editor agreement | More than 30 editors and advisors resign; no successor | (a), policy change | Public |
| Review of Social Economy | Taylor & Francis | 2026 | Owning association renews publisher contract | 41 editors and board members resign; new journal planned | (c), society-owned | Public |
| Games and Economic Behavior | Elsevier | 2026 | Editor-in-chief not renewed after under six years | Editors and 58 of 62 advisory editors resign; society ends association, exploring successor | (c) in progress, high $\theta$ | Public |

Note: Model cases from Proposition 1: (a) the owner's change of editor or policy proceeds and the board stays or disperses without a successor; (c) the change proceeds and the board exits with a successor. Sources: Chronicle of Higher Education (2003); Takagi (2020); Southern Illinois University (2007); ACAES (n.d.); Wiemer (2020); MDPI (2024); Game Theory Society (2026); Retraction Watch (2024, 2026a, 2026b, 2026c); for FRL, the author's participation (Section 7).

### *6.1 The episodes in brief*

**European Economic Review.** The European Economic Association had adopted the EER, published by Elsevier since 1969, as its official journal. In the early 2000s, a dispute over Elsevier's pricing led the association to cut its ties with the journal; Theodore Bergstrom, whose work documented the price gap between commercial and non-profit journals (Bergstrom, 2001), had included the EER in a "rogues' gallery" of overpriced journals (Chronicle of Higher Education, 2003). The association replaced the EER with a new official journal, the JEEA, launched with MIT Press in 2003. Two former EER editors were to move to the new journal once the noncompete clauses in their Elsevier contracts expired (Chronicle of Higher Education, 2003). This is the clearest case of successful exit: the society, not the title, carried the reputation.

**Journal of Banking & Finance.** Giorgio Szegö co-founded the JBF in 1977 and led it for about 30 years (D'Ecclesia, Carulli Szegö, and Zambruno, 2023). In April 2007, Southern Illinois University announced that Ike Mathur would become managing editor from January 1, 2008, following Szegö after 31 years (Southern Illinois University, 2007). A former associate editor later wrote that Elsevier

"removed the editor and moved the location of the journal from Europe to the United States," and that the journal, once closely associated with a group of European scholars, became "only one of many faceless journals competing for attention from cyberspace" (Takagi, 2020).

**Journal of Asian Economics.** On March 18, 2020, Elsevier offered the editor-in-chief, Calla Wiemer, a one-year terminal contract for 2020, signalling that the publisher would take over the selection of editors and ending the role of ACAES, which had chosen each of the journal's three editors-in-chief since 1990. Stakeholders, including a former editor-in-chief and a current executive editor, wrote to Elsevier in support of Wiemer. ACAES records that Elsevier "stood by its decision to terminate the ACAES Editor-in-Chief and, with that, the relationship to ACAES" (ACAES, n.d.; Wiemer, 2020). The journal is now managed by Elsevier Beijing, which appoints the editor.

**Games and Economic Behavior.** Described in the introduction. Moulin was the second editor in the journal's history, after the founding editor Ehud Kalai, who had led it for 31 years (Game Theory Society, 2020). The resigning editors gave three months' notice and committed to handling submissions until November 14, 2026 (Retraction Watch, 2026a). On October 9, 2026, the Game Theory Society ended its association with the journal from November 15 and said it was exploring a society-owned open-access journal (Game Theory Society, 2026).

**Non-Elsevier episodes.** At *Econometrics* (MDPI), the editors and the publisher "were unable to agree on an overarching strategy for the development of this journal" (MDPI, 2024). At the *Journal of Economic Surveys* (Wiley), more than 30 editors and advisors resigned over what they saw as an emphasis on quantity over quality and the terms of a new editor agreement (Retraction Watch, 2024). At the *Review of Social Economy* (Taylor & Francis), 41 editors, associate editors and advisory board members resigned after the owning association renewed its publishing contract, and said they planned to launch their own journal (Retraction Watch, 2026b). None of these involved the publisher removing an editor.

## 7. The Finance Research Letters transition

### *7.1 Disclosure*

I was one of the three editors appointed by Elsevier to take over FRL from January 1, 2015, with Brian Lucey and Jonathan Batten, and I served as co-editor-in-chief until 2017. I co-wrote and co-signed the incoming editors' note to the board described below and helped recruit new board members. At the time, I accepted the account that the outgoing editor had been given ample notice and bore responsibility for the lack of communication with the board; I now regard that account as incomplete, because, as Section 7.2 records, the board was told of the change only after the decision and the appointments had been made. The growth in FRL's output documented in Section 9 and in Cumming (2026a) began during, and continued after, my tenure. I have had a dispute with Elsevier arising from a different matter, and I have co-authored with one of the other incoming editors. The account below relies on public records, on facts I observed as a participant, and on the incoming editors' own note.

### *7.2 The transition through the lens of the model*

FRL was founded in 2004 and edited from its launch by Ramo Gençay. In 2014, Elsevier declined to renew his contract. As an incoming editor, I was told that the ground was a publisher policy limiting chief editors to ten years. Elsevier selected and appointed the three incoming editors, and the board was informed of the change in November 2014, after the decision and the appointments had been made. Several board members resigned in the following days, citing the lack of consultation, and others questioned the process; no organization was in a position to host a successor, and none was launched.

The incoming editors' note to the board set out the plan for the journal. Its central objective was to greatly increase the flow of submissions, which it described as a high priority for Elsevier and for the incoming editors. It proposed a review model in which associate editors would serve as the primary referees, with an "up or out" approach under which papers with shortcomings that could be described briefly would be conditionally accepted and others rejected; a broader scope; links with major finance conferences; and a "reduce, revise, resubmission" option to be developed with editors of other Elsevier journals.

In the model's terms, the episode is case (a) of Proposition 1. The owner's gain $G$ lay in volume, the objective at the head of the plan. The board's share of the reputation, $\theta$, was low: FRL was not the official journal of any society. And the cost of starting again, $F$, was high: in finance in 2014 there was no established non-profit host for a letters journal. The model therefore predicts that the owner proceeds, that board members leave individually rather than collectively, that no successor is founded, and that the owner's policy is implemented. The first three are what happened. The fourth is tested in Section 9.

## 8. P3: the ten-year cap and its enforcement

Proposition 3 distinguishes a binding cap on tenure from one enforced at the owner's discretion. I test it with a dataset of chief and lead editors at 65 Elsevier economics, finance, and accounting journals and 45 journals at other publishers and societies, compiled from public sources: editors' curricula vitae, society and university announcements, journal editorials, and RePEc listings (Section 4.4). The test sample consists of editors who had at least ten years of tenure at some point between 2010 and 2015 while in post, the period to which the cap applied; in practice this means spells that began in 2005 or earlier. Both the start and the end of the spell, or the editor's continuation in post beyond 2015, can be verified for 26 such editors at Elsevier journals and 10 elsewhere. Table 6 lists them.

**Table 6: Long-serving editors during 2010–2015 and whether they left**

| Journal | Editor | Start | End | Tenure in 2010 | ABS 2010 | Left 2010–15 | Stated reason |
|---|---|---|---|---|---|---|---|
| **Panel A: Elsevier journals** | | | | | | | |
| Economic Modelling | P. Pauly | 1997 | 2015 | 13 | 2 | Yes | Co-editor; reason not stated |
| Economic Systems | R. Frensch | 2001 | c. 2025 | 9 | 1 | No | |
| Economics Letters | E. S. Maskin | 1992 | 2011 | 18 | 3 | Yes | Became advisory editor |
| Economics of Education Review | E. Cohn | 1981 | c. 2013 | 29 | 2 | Yes | Founding editor; reason not stated |
| Energy Economics | R. S. J. Tol | 2004 | 2025 | 6 | 2 | No | |
| Finance Research Letters | R. Gençay | 2004 | 2014 | 6 | n.r. | Yes | Not renewed; ten-year cap cited |
| Games and Economic Behavior | E. Kalai | 1989 | 2020 | 21 | 4 | No | |
| Global Finance Journal | M. Shahrokhi | 1989 | 2016 | 21 | 2 | No | |
| International Review of Economics & Finance | H. Beladi | 1992 | 2018 | 18 | n.r. | No | |
| International Review of Economics & Finance | C. R. Chen | 1992 | 2022 | 18 | n.r. | No | |
| International Review of Law and Economics | A. W. Katz | 2003 | 2013 | 7 | 2 | Yes | Co-editor; reason not stated |
| Journal of Accounting and Economics | S. P. Kothari | 1997 | 2019 | 13 | 4 | No | |
| Journal of Accounting and Economics | R. W. Holthausen | 2005 | 2022 | 5 | 4 | No | |
| Journal of Accounting and Public Policy | L. A. Gordon | 2001 | 2018 | 9 | n.r. | No | |
| Journal of Corporate Finance | J. M. Netter | 2001 | 2017 | 9 | 3 | No | |
| Journal of Corporate Finance | A. B. Poulsen | 2001 | 2017 | 9 | 3 | No | |
| Journal of Economic Dynamics and Control | C. H. Hommes | 2002 | 2012 | 8 | 3 | Yes | Co-editor; reason not stated |
| Journal of Economic Theory | K. Shell | 1968 | 2019 | 42 | 4 | No | |
| Journal of Economics and Business | K. J. Kopecky | 1993 | 2017 or later | 17 | 1 | No | |

| Journal | Editor | Start | End | Tenure in 2010 | ABS 2010 | Left 2010–15 | Stated reason |
|---|---|---|---|---|---|---|---|
| Journal of Financial Economics | G. W. Schwert | 1995 | 2021 | 15 | 4 | No | |
| Journal of Financial Stability | I. Hasan | 2004 | In post | 6 | n.r. | No | |
| Journal of Health Economics | J. P. Newhouse | 1981 | 2011 | 29 | 3 | Yes | Founding editor; reason not stated |
| Journal of International Economics | C. Engel | 2001 | 2017 | 9 | 3 | No | |
| Journal of Macroeconomics | T. Palivos | 2000 | 2013 | 10 | 2 | Yes | Co-editor; reason not stated |
| Pacific-Basin Finance Journal | S. G. Rhee | 1992 | 2017 | 18 | 2 | No | |
| The International Journal of Accounting | A. R. Abdel-khalik | 2000 | In post 2025 | 10 | 3 | No | |
| **Panel B: Other publishers and societies** | | | | | | | |
| Economic Journal | S. Machin | 1998 | 2013 | 12 | 4 | Yes | |
| Journal of Financial and Quantitative Analysis | P. Malatesta | 1987 | 2022 | 23 | 4 | No | |
| Journal of Law and Economics | S. Peltzman | 1991 | In post | 19 | n.r. | No | |
| Journal of Money, Credit and Banking | K. D. West | 2001 | In post | 9 | 4 | No | |
| Journal of Risk and Uncertainty | W. K. Viscusi | 1988 | In post | 22 | 4 | No | |
| Public Choice | W. F. Shughart II | 2005 | 2022 | 5 | 3 | No | |
| Quarterly Journal of Economics | L. F. Katz | 1991 | In post | 19 | 4 | No | |
| Quarterly Journal of Economics | R. J. Barro | 2004 | In post | 6 | 4 | No | |
| Review of Quantitative Finance and Accounting | C.-F. Lee | 1990 | In post | 20 | 3 | No | |
| Small Business Economics | D. B. Audretsch | 1989 | In post | 21 | 3 | No | |
| **Tests** | | | | | | | |
| Elsevier editors who left in 2010–2015 | 8 of 26 | | | | | | |
| Binomial test, enforcement probability 0.9 / 0.75 / 0.5 | $p < 10^{-12}$ / $p < 10^{-5}$ / $p = 0.04$ | | | | | | |
| Other editors who left in 2010–2015 | 1 of 10 | | | | | | |
| Fisher exact test, Elsevier versus others (one-sided) | $p = 0.20$ | | | | | | |
| Elsevier editors who left, by ABS 2010 grade 4 / 3 / 2 / 1 | 0 of 5 / 3 of 7 / 4 of 7 / 0 of 2 | | | | | | |
| Fisher exact test, grade 4 versus below (one-sided) | $p = 0.10$ | | | | | | |
| Rank-sum (Mann–Whitney) test, grades of leavers versus stayers (one-sided, normal approximation) | $p = 0.14$ | | | | | | |

Note: Editors in the top editorial position (editor-in-chief, managing editor, or co-equal lead editor) whose spells began before 2014 and who had at least ten years of tenure at some point in 2010–2015 while in post, with start and end years verified from public sources. “c.” marks years bounded by a successor’s documented start. ABS 2010 is the journal’s grade in the Association of Business Schools Academic Journal Quality Guide, version 4 (March 2010), from 1 to 4, which predates the period of the test; n.r. means not rated. The tests by grade use the 21 Elsevier editors at rated journals. A departure with no stated reason is not presumed to be a non-renewal, and no inference is drawn about the reasons for any individual departure. One-sided binomial tests of the hypothesis that each such Elsevier

editor left by 2015 with the stated probability. The comparison panel was not extended in the second round of data collection. Source: author's tenure dataset (Appendix B).

A binding cap is rejected decisively. Of the 26 Elsevier editors, eight left between 2010 and 2015 and 18 stayed on, in most cases for many more years (Figure 3). Under a binding cap every line in the figure would end at its ten-year tick or by 2015; instead most run well past both; those who stayed include the managing editor of the *Journal of Financial Economics* and the founding editors of GEB, the *Journal of Economic Theory*, and the *Pacific-Basin Finance Journal*. Even if the cap had been enforced with probability 0.9, observing eight or fewer departures among 26 editors would have a probability below $10^{-12}$; even at probability 0.5 it would be 0.04. Treating each journal rather than each editor as the unit, because co-editors at the same journal face one decision, eight of the 23 journals lost a long-serving editor, which has a probability below $10^{-9}$ under enforcement with probability 0.9 and of 0.11 at 0.5. Long-serving Elsevier editors left at a higher rate than long-serving editors at other publishers, where no such cap applied (31 against 10 percent), which is the direction a partly enforced cap would imply, but the difference is not significant (8 of 26 against 1 of 10; Fisher exact $p = 0.20$, one-sided), although the comparison panel is smaller and was not extended in the second round of data collection.

The reasons for the eight departures bear on the second part of Proposition 3. Only one is documented as a non-renewal: FRL's founding editor in 2014, the case in which the cap was cited. One editor became an advisory editor; two were founding editors who left after about 30 years; and four were co-editors in multi-editor teams, for whom no reason is recorded. The one documented enforcement occurred at a journal that was weak on most measures, as Proposition 3(b) predicts, though not uniquely so. Among the seven Elsevier finance journals in Panel A with SCImago data, FRL had the lowest estimated revenue in 2013 (Section 3.5) and the third-lowest SCImago Journal Rank (0.49, above the *Journal of Economics and Business* at 0.39 and the *Pacific-Basin Finance Journal* at 0.46, and below the *International Review of Economics & Finance* at 0.72, the *Journal of Financial Stability* at 1.70, the *Journal of Corporate Finance* at 1.91, and the *Journal of Financial Economics* at 12.3). It was not rated in the 2010 ABS guide, and it had neither a society nor a long publication record. The long-serving editors of the two journals ranked below it by SCImago both stayed. The second round of data collection, which nearly doubled the Elsevier sample, found no other departure attributed to the cap. The test therefore establishes Proposition 3(a), that the cap did not bind, with confidence, while the evidence for 3(b), that enforcement falls on weak boards, still rests on a single documented event. Distinguishing replacement from voluntary departure in the remaining cases would require renewal records that are not public.

A ranking fixed before the period of the test gives a broader, if weaker, check. The ABS Academic Journal Quality Guide of March 2010 grades journals from 1 to 4, and it rates 21 of the 26 Elsevier journals in Table 6 (Association of Business Schools, 2010). None of the five editors at journals graded 4, GEB, the *Journal of Accounting and Economics*, the *Journal of Economic Theory*, and the *Journal of Financial Economics*, left in 2010–2015, against 7 of the 16 at journals graded 3 or below (Figure 3). FRL, founded in 2004, was not yet rated. The pattern runs in the direction Proposition 3(b) predicts, in that departures were absent where the board carries most of the reputation, but it is not statistically significant (Fisher exact $p = 0.10$, one-sided; Mann–Whitney $p = 0.14$), departures do not fall steadily with the grade (neither of the two editors at journals graded 1 left), and since the seven departures at rated journals have no stated reason, they cannot be read as enforcement. The ranking evidence is therefore consistent with selective enforcement but does not establish it.

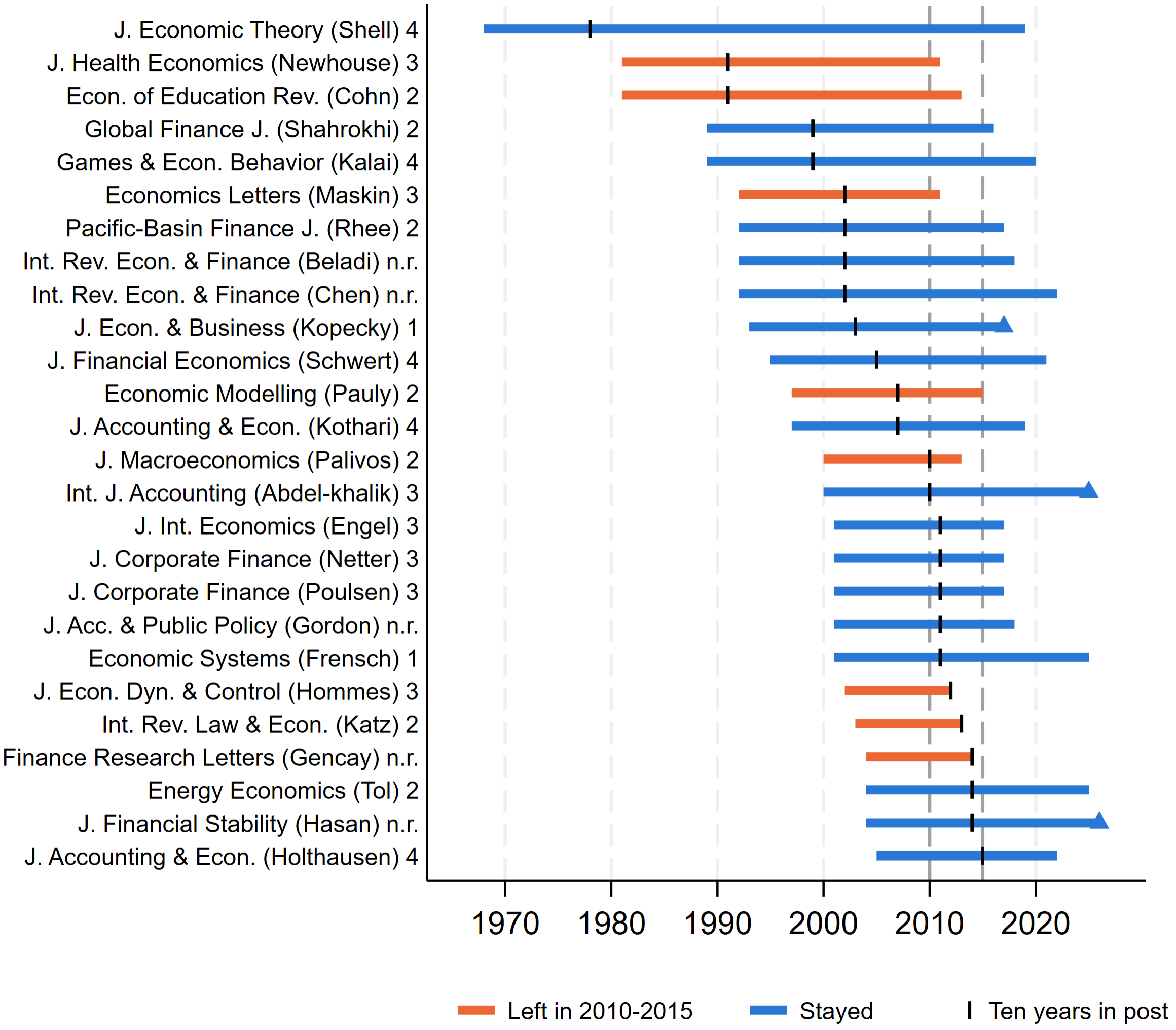

**Figure 3: Tenure of the 26 long-serving Elsevier editors and the ten-year mark**

Note: Each line is one editor's spell in the top editorial position, from start to end; triangles mark editors still in post at the last verified date. Orange lines are editors who left in 2010–2015, blue lines editors who stayed. The black tick marks the editor's tenth year in post, and the dashed vertical lines mark the 2010–2015 test window. The number after each name is the journal's grade in the 2010 ABS guide (n.r. = not rated). Under a binding ten-year cap, every line would end at its tick or by 2015. Source: Table 6.

An alternative explanation is that the owner enforces the cap where a journal earns little and leaves it alone where the journal makes money, whatever the strength of the board. This explanation is not outside the model. The owner's loss from exit, $L(\tau) = \theta(\tau)mR$, rises with the journal's revenue $R$, and its gain from a change of policy, $G$, is largest where a title has room to grow, so Proposition 3(b) predicts enforcement at low-revenue journals with room to expand as well as at journals with weak boards. The data cannot separate the two channels, because the one documented enforcement scores lowest on both. On the estimates of Cumming (2026a), FRL had the lowest revenue of the Elsevier finance journals in Table 6 in 2013 ($0.37 million), its output was falling, and its letters format suited expansion; it was also among the lowest-ranked of them and had no society. Its estimated revenue then rose to $1.9 million by 2018, the end of the window in Section 9, and to $16.5 million by 2024, a later path that also reflects the journal network launched after 2018 (Cumming, 2026a). The *Journal of Economics and Business*, also low in revenue ($0.46 million in 2013) and graded 1 by ABS, kept its long-serving editor, which fits the revenue channel only if its owner saw little room for growth. Separating revenue from board strength would require enforcement events at journals that are profitable but weakly governed, or unprofitable but strongly governed, and the sample contains none. The safer reading is that the owner applied the cap where its expected gain $G$ was high relative to its expected loss $H(F*)L$, which is what Proposition 3(b) says, without a claim about which term did the work.

Replacements of long-serving editors also occur outside the test window, and are not counted in it. On 25 July 2025, Elsevier announced that Boqiang Lin had succeeded Richard Tol, who had led *Energy Economics* since 2004, as editor-in-chief (Elsevier, 2025); Tol has written that Elsevier ended his appointment with immediate effect on 17 July 2025, and has given his own account of the reasons (Tol, 2025).

Two further caveats apply. The sample is limited to editors whose tenure can be verified from public sources, which favours journals with documented histories; editors at less prominent Elsevier journals, where the model predicts enforcement would be more likely, remain under-represented. And because the reasons for seven of the eight departures are not documented, eight is an upper bound on enforcement in this sample, and one a lower bound.

## 9. P4: journal-level consequences

Proposition 1(a) predicts that where a replacement proceeds and the board disperses, the owner's policy is implemented (P4). Where the owner's gain lies in volume, as the FRL plan in Section 7 indicates, publication volume should rise relative to comparable journals, and citations per document need not. The prediction can be tested with journal-level bibliometric data. I use the SCImago panel of 26 finance journals from Cumming (2026a) and three outcomes: the log of citable documents, which captures volume; two-year citations per document, a coarse measure of standing; and the share of internationally co-authored papers, a measure of author composition. The panel covers finance journals, so I estimate the design for the JBF (event year 2008, the first year under the new managing editor) and FRL (2015). The JAE requires an economics panel and GEB requires post-event data; both are left for future work.

For each event, I estimate a two-way fixed-effects difference-in-differences that compares the treated journal with the finance journals outside the Elsevier Finance Journal Ecosystem (15 controls for FRL and 14 for the JBF):

$$y(\mathrm{j,t}) = \alpha(\mathrm{j}) + \lambda(\mathrm{t}) + \beta \cdot \mathit{Treated}(\mathrm{j}) \cdot \mathit{Post}(\mathrm{t}) + \varepsilon(\mathrm{j,t})$$

The FRL window is 2010–2018, ending before the ecosystem was launched, and the JBF window is 2003–2014. With a single treated journal per event, conventional standard errors are not informative, so inference uses placebo-in-space: I re-estimate β assigning the treatment to each control journal in turn and report the share of placebo estimates at least as large in absolute value as the actual estimate. For FRL I also construct a synthetic control for log documents (Abadie, Diamond, and Hainmueller, 2010), choosing non-negative donor weights that sum to one to match FRL's pre-2015 path.

**Table 7: Journal outcomes after editor replacement at the JBF (2008) and FRL (2015)**

| Event and outcome | DiD estimate (β) | Placebo p | Controls | Treated mean, pre | Treated mean, post |
|---|---|---|---|---|---|
| **FRL (2015), window 2010–2018** | | | | | |
| log(documents) | 1.334 | 0.06 | 15 | 3.39 | 4.95 |
| Citations per document (2-yr) | 0.477 | 0.25 | 15 | 0.85 | 1.32 |
| International co-authorship (%) | 2.94 | 0.38 | 15 | 25.4 | 33.5 |
| log(estimated revenue) | 0.751 | 0.13 | 15 | 13.24 | 14.20 |
| **JBF (2008), window 2003–2014** | | | | | |
| log(documents) | 0.244 | 0.53 | 14 | 4.98 | 5.59 |
| Citations per document (2-yr) | 1.103 | 0.13 | 14 | 1.86 | 3.92 |
| International co-authorship (%) | −2.09 | 0.67 | 14 | 29.8 | 35.7 |
| log(estimated revenue) | 0.180 | 0.67 | 14 | 15.01 | 15.53 |

Note: Two-way (journal and year) fixed-effects difference-in-differences of each outcome on Treated × Post, one treated journal per event. Controls are the non-ecosystem finance journals in Cumming (2026a), excluding the Journal of Commodity Markets (founded 2016). Placebo p = (1 + number of control journals whose placebo estimate is at least as large in absolute value) / (1 + number of controls). Synthetic control for FRL log documents: weights 0.68 on the Journal of Financial Markets and 0.32 on the Quarterly Review of Economics and Finance; pre-2015 root mean squared prediction error 0.16; post-2015 gaps of 1.17, 1.43, 1.25 and 1.21 log points in 2015–2018. Estimated revenue is the journal-level revenue estimate in Cumming (2026a), in US dollars. Source: SCImago, via the replication package of Cumming (2026a).

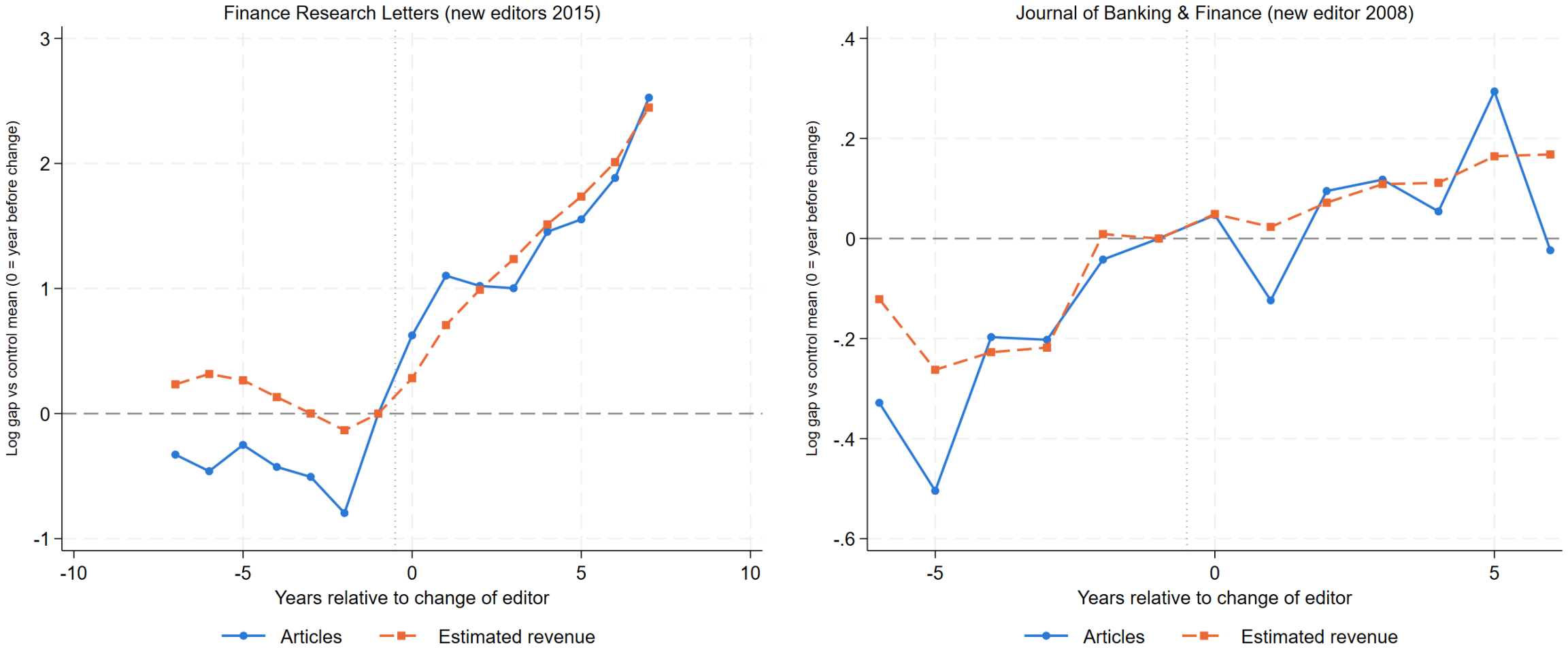

**Figure 4: Output and estimated revenue relative to control journals, by year relative to the change of editor**

Note: Difference between the treated journal's log value and the mean log value of the non-ecosystem control journals, normalized to zero in the year before the change. FRL is shown for 2008–2022 and the JBF for 2002–2014; the estimates in Table 7 use the windows 2010–2018 and 2003–2014. Source: SCImago and revenue estimates from Cumming (2026a).

At FRL, log documents rose by 1.33 relative to the controls after 2015, roughly a 3.8-fold relative increase. This is the largest estimate among FRL and its 15 placebo assignments (placebo $p = 0.06$, the smallest value attainable with 15 controls). The synthetic control tracks FRL closely before 2015, with a pre-period prediction error of 0.16 log points, and FRL then exceeds it by 1.2 to 1.4 log points in every year from 2015 to 2018. Citations per document and international co-authorship did not change significantly. Estimated revenue rose by 0.75 log points relative to the controls, about 2.1-fold, the second largest of the 16 assignments (placebo $p = 0.13$); in dollars, from \$0.37 million in 2013 to \$1.86 million in 2018. Figure 4 shows both paths. Two caveats apply. FRL's output had already doubled in 2014, from 23 to 52 documents, the last year under the outgoing editor. This may reflect a policy that predated the change of editor; in the model what matters is the owner's policy, so the 2014 rise is consistent with P4, but it makes 2015 a less sharp treatment date, and estimates measured relative to 2014 understate the change relative to 2013; and the window ends in 2018 to exclude the ecosystem, which brought a far larger expansion afterwards (Cumming, 2026a). At the JBF, by contrast, there is no detectable change in volume after 2008 (0.24, $p = 0.53$) or in estimated revenue (0.18, $p = 0.67$). Citations per document rose temporarily in 2009–2011, but the estimate is not significant (1.10, $p = 0.13$), and international co-authorship did not change. The JBF is also a weak test of P4, since the model predicts movement only toward the owner's objective and none is documented there; a null result is consistent with the model but cannot confirm it. The JBF result should be read with care for a further reason: it was the largest journal in the panel, with 194 documents in 2007 against 113 for the next largest, and no weighted combination of controls fits its pre-2008 path well.

Taken together, the economic consequences of an owner-initiated change of editor depend on whether the board leaves with somewhere to go. Where it disperses, as at FRL, the owner captures the short-run gain from the new policy; where nothing changes in the policy, as at the JBF, neither output nor revenue moves; and where a society leaves with the board, as at the EER, the owner keeps the title but loses standing to the successor (Section 3.5). The evidence therefore supports the volume prediction where the operating plan was explicitly about volume, at FRL, and not where the documented change was the departure of a long-serving editor and the relocation of the editorial office, at the JBF. Two events cannot establish a general pattern; the same design can be applied to the editor-replacement episodes in Appendix A as journal-level data are assembled for other fields.

### *9.1 Short-run gains and long-run costs*

Growth in output and revenue measures the owner's gain, not the journal's welfare, and the two can diverge. The model takes the reputation $R$ as given once the board has dispersed. If the owner's policy itself draws the reputation down, for example because rapid growth relies on practices that raise measured citations without raising the quality or influence of the work, the owner's long-run payoff is $G$ less the eventual fall in $mR$, and a short window observes only $G$. Alexander and Cumming (2026) examine this directly for the same 26 finance journals as the panel used in Table 7. Hierarchical clustering of their citation metrics places FRL in a small "quickly-evolving" cluster of eight journals whose publication volume and citations grew almost exponentially after the ecosystem was formed in 2019. Within that cluster, a small, tightly networked group of highly prolific authors accounts for a large share of the citation growth: the four most-cited authors in the cluster made 1,006 self-citations and 803 citations to each other's work in 2019–2024, against 90 and 53 for the four most-cited authors in the three leading finance journals. The growth was not matched by quality-adjusted measures: the SCImago Journal Rank of the cluster's journals showed no comparable increase, while submission-to-acceptance times fell. In their data, FRL has the second-highest two-year impact factor of the 26 journals, behind only the *Journal of Financial Economics*, and the shortest time from submission to acceptance reported on the journals' websites, 101 days. Alexander and Cumming treat the ecosystem as context rather than cause and do not attribute intent to any editor or author. Their evidence nonetheless suggests that the growth in Table 7 and Figure 4 raised measured impact more than standing, and is better read as a rise in measured impact than as evidence of higher standing. This is the distortion that multitask agency theory predicts when rewards are tied to what can be measured, here volume and citations, rather than to the harder-to-measure quality of the work (Holmström and Milgrom, 1991).

Reputational losses of this kind arrive slowly, which is why a short window can make a volume strategy look profitable. Indexing services withdraw impact factors only after a citation pattern has persisted (Retraction Watch, 2025), and ranking lists are revised every few years. Alexander and Cumming (2026) argue that a divergence between raw citation counts and quality-adjusted metrics is the signal that ranking bodies such as the Chartered Association of Business Schools should monitor. Because the lists move slowly, the stickiness that slows a successor's entry onto the lists (Section 3.5) also delays the downgrade of a title whose standing falls. In the model, such lags weaken the incentive to protect the reputation, because the cost of a policy arrives long after its gain; Tadelis (1999) shows more generally that a name can be a tradeable asset whose value its owner can choose to maintain or run down. The history observed here ends in 2026, about a decade after the FRL transition and seven years after the network launched. The estimates are therefore short-run effects on volume and revenue. They do not show whether the journal's standing, its value to authors, or the value of the title to its owner will be higher or lower once the lagged consequences, such as ranking revisions, changes in indexing, or authors' responses, have occurred.

## 10. The predictions and the evidence

Table 8 brings the evidence together. P1 is supported: in none of the 19 editor-replacement episodes did an owner reverse course after the board resigned, the three reversals after resignation all concerned individual papers, as Proposition 6 predicts, and the documented withdrawal of an appointment followed collective protest and came before resignations. P2 is largely supported: every comparison of the counterparts of $\theta$ and $F$ has the predicted sign, and society ownership and medicine and the life sciences are significant whether or not announced successors are counted; the society-affiliation estimate is large but depends on two announced successors, and editor replacement, mapped to a high cost of staying, has the opposite sign to the one the model implies. On P3, a binding cap is rejected decisively, while the evidence that enforcement targets weak boards rests on a single documented event and a pattern by journal grade that has the predicted direction but is not significant. P4 is supported for FRL, where the operating plan was about volume, and the absence of a volume change at the JBF, where no volume objective is documented, is consistent with the model. P5 is not supported in its simplest form, although selection means the model does not require the share of exits to rise.

**Table 8: The model's predictions and the evidence**

| Prediction | Evidence | Assessment |
|---|---|---|
| P1. Reversals precede resignations | No reversal in any of 19 editor-replacement episodes after resignation; the three reversals after resignation all concerned individual papers; documented withdrawal (European Law Journal, 2018) followed collective protest and came before the board resigned (Section 5.2) | Supported |
| P2. Exit depends on the outside option | Society affiliation +43 points, society ownership −24 points, medicine and life sciences −33 points in the probability of a successor launched or announced (Table 4); with launched successors only, +18 (not significant), −37 and −26 points (Section 5.3) | Supported for society ownership and field; society affiliation depends on announced successors |
| P3. Tenure cap not binding; enforced against weak boards | 8 of 26 long-serving Elsevier editors left in 2010–2015 (31 percent, against 10 percent elsewhere; $p = 0.20$); binding cap rejected ($p < 10^{-12}$ at 0.9 enforcement); the only documented non-renewal was at the lowest-revenue journal, which was unrated by ABS; no editor at an ABS grade-4 journal left, against 7 of 16 below ($p = 0.10$) (Table 6) | Non-binding: supported. Selectivity: consistent in direction, not significant |
| P4. Owner's policy implemented when the board disperses | FRL volume about 3.8-fold relative to controls after 2015, citations unchanged; no volume change at JBF after 2008 (Table 7) | Supported for FRL; JBF uninformative |
| P5. Cheaper hosting raises exit | Successor share 44% before and after 2023; non-commercial hosts' share of successors rose from 5 of 8 to 13 of 15 (Section 5.4) | No rise in share, which the model does not rule out; consistent in hosts |

Note: Predictions from Section 2. Section and table references are to the evidence reported in this paper.

The GEB episode in 2026 combines the features the model associates with a successful exit, a collective departure by a board whose reputation is tied to a society that has now ended its association with the title and is exploring a successor of its own. On the model and the evidence, it is the episode most likely to follow the path of the European Economic Review two decades ago, with the reputation moving to a new title while the old title continues under new editors.

## 11. Discussion

### *11.1 Caveats*

Six caveats apply. First, the Retraction Watch list is not a census: it depends on press coverage and favours episodes that became public, so the counts in Section 5 are lower bounds and should not be compared across publishers of different sizes. Second, many codes, especially ownership, rest on inference from public sources; the dataset flags each inference, and the ownership result is directionally robust but imprecise when inferred codes are dropped. Third, the samples are small: the estimates in Tables 3 and 4 should be read as descriptive, the tenure test in Table 6 rests on editors whose histories are documented publicly, and the journal-level estimates in Table 7 rest on two events. Fourth, I was a participant in the FRL transition and have had a dispute with the publisher arising from a different matter; I have tried to address this by relying on public records and on facts I observed, by disclosing my own view at the time, and by grounding the main tests in public data on all publishers. Fifth, the empirical counterparts are imperfect: a society can both raise $\theta$ and lower $F$, so the ownership comparisons do not separate the two, and the field comparison may reflect differences in indexing and norms as well as in hosting. Sixth, a replacement can be good for a journal. Nothing in this paper shows that any particular editorial decision was wrong; the question is who decides, on what stated grounds, and what happens when the board objects.

### *11.2 Implications for research policy*

The findings have implications for five groups. *Editorial boards*: because protest succeeds only before resignations (P1) and works best when it is public and collective (Proposition 5), boards have most influence if they hold a consultation right over the appointment of a chief editor, exercised before a decision is announced. *Scholarly societies*: because society affiliation raises the board's outside option (P2), societies that designate a publisher-owned journal as their official journal can secure, in contract, the right to appoint or approve the editor, or ownership of the title; the *Journal of Asian Economics* shows the cost of relying on convention. *Funders and libraries*: because successors follow where non-profit hosts exist and are rare where they do not (P2 and P5), support for diamond open-access infrastructure, of the kind cOAlition S (2018) has promoted, lowers the cost of starting again and so strengthens the bargaining position of editorial communities in every field, not only those that already have hosts. *Ranking bodies and evaluation committees*: because ranking lists attach reputation to titles (Section 3.5), the bodies that compile them could treat an owner-initiated replacement followed by collective board exit as information about a journal, extend provisional ratings to successors founded by most of a departing board, and monitor the divergence between raw citation counts and quality-adjusted measures that Alexander and Cumming (2026) document, in line with the San Francisco Declaration on Research Assessment (DORA, 2012) and the Leiden Manifesto (Hicks et al., 2015). *Publishers*: because discretionary rules are enforced selectively (P3), publishing policies on editor terms and appointments, and the grounds for non-renewal, would let boards and authors see whether those policies are applied consistently.

In the model, each of these measures either raises the share of the reputation that a departing board can take or lowers its cost of starting again, which makes owners' threats less credible and boards' voice more effective. They do not require any change in who owns journals; they change the institutions around ownership that decide whether editorial communities can leave.

### *11.3 Beyond academic journals*

The setting is narrow, but its structure is common. In professional-service firms, software projects, newsrooms, and technology start-ups, an owner holds an alienable name, brand, or legal entity, while much of the value rests on people who can leave. Four lessons from journals carry over.

First, whether the people who carry the value can leave with it depends less on them than on institutions outside the organization that can receive them. After Oracle acquired the OpenOffice.org office suite with Sun Microsystems, the project's volunteer developers formed The Document Foundation in 2010 to continue the software as LibreOffice; the brand they had built stayed with Oracle, which they invited to donate it (LWN.net, 2010). When the owner of the Terraform infrastructure software moved it to a more restrictive license in 2023, the Linux Foundation took in the community's fork, OpenTofu, under neutral governance (Linux Foundation, 2023). After the owners of the sports site Deadspin told its writers in 2019 to "stick to sports" and the editorial staff resigned, 18 of them founded Defector as an employee-owned company (Defector, n.d.). In each case a foundation or a cooperative played the role a society plays for a journal: it lowered the cost of starting again, $F$, and helped the group leave together, while the name stayed with the owner.

Second, voice works when exit is credible, and it works before exit, not after. In November 2023, after the board of OpenAI removed its chief executive, about 600 of its roughly 770 employees signed a letter threatening to leave for Microsoft, which had said it had positions for all of them (Data Center Knowledge, 2023); within days the chief executive was reinstated and the board replaced (NBC Chicago, 2023). In the model's terms, a rival able to absorb the whole team made the threat credible, and the decision was reversed before anyone left, the case (b) that a list of resignations rarely records. The same reasoning applies to partners who can take clients from law, consulting, and accounting firms, to physician groups in hospital systems, and to research teams in companies. Rules that restrict departure, such as noncompete agreements, raise $F$ and so weaken voice as well as exit (Marx, Strumsky, and Fleming, 2009; Garmaise, 2011).

Third, third-party ratings that attach to names rather than people move reputation toward owners. Restaurant guides rate restaurants rather than chefs, league tables and legal directories rank firms, and credit ratings attach to issuers. Like journal ranking lists, they make part of the value of human capital alienable, raise the cost of leaving, and delay the penalty when the quality behind a name declines. Rating bodies that let a rating follow the people who earned it, or that give provisional ratings to successors founded by departing teams, would shift bargaining power back toward those people.

Fourth, rules that owners announce, such as term limits for chief executives and directors, mandatory retirement ages, and rotation requirements, constrain only if they bind. The test in Section 8 can be applied wherever such a rule exists: compare the share of those reaching the limit who leave with the share a binding rule implies, and ask whether departures are concentrated where the people affected have weak outside options. A broader caution follows from Section 9.1. Where reputational costs arrive with long lags, the gains from overriding the people who carry a reputation appear first, and the costs may not appear within any short window of data, in journals or elsewhere.

## 12. Conclusion

Academic journals separate an alienable asset, the title, from inalienable human capital, the reputation embodied in editors and boards, and they make disputes between the two observable. A simple property-rights model of that separation predicts when owners will replace editors, when protest will succeed, and when boards will leave and take the reputation with them. The evidence from 57 recorded mass resignations and from economics and finance supports the model's central predictions: owners back down, when they do, before boards resign; exit succeeds where a society or a non-commercial host gives the board somewhere to go, and ranking lists that rate titles rather than boards make it harder; a stated cap on tenure did not bind and was enforced at the least lucrative journal in the comparison; and where boards disperse, the owner's policy prevails, as in the almost fourfold growth of FRL's output relative to comparable journals after 2015. That growth is a short-run gain to the owner, not evidence that the change was good for the journal: the reputational costs of a volume strategy arrive with long lags, and the history observed here is truncated at 2026. For research policy, the lesson is that the governance of journals is set less by publishers' formal rights than by the institutions, societies, non-profit hosts, funders and ranking lists, that determine whether editorial communities can leave with a journal's standing. The GEB episode will show whether a collective exit backed by a society can do in game theory what the European Economic Association did in economics two decades ago. The same questions, about who holds the name, who carries the value, and which institutions let the value move, arise wherever an owner's title and its people's reputation can part company, from software projects and newsrooms to professional partnerships and technology firms (Section 11.3).

## Appendix A. Coded Retraction Watch episodes

The table reports the main codes for each episode on the Retraction Watch Mass Resignations List (Retraction Watch, 2026c). Owner = owner of the title at the time of the dispute. Soc. = 1 if the journal was owned by, or was the official journal of, a scholarly society. Ed. repl. = 1 if the publisher or owner removed, declined to renew, or replaced the editors. Successor = successor journal launched or planned by the departing editors or society. Rev. = 1 if the owner reversed or materially changed the disputed decision. Secondary triggers, successor hosts, sources for each code, and notes on inferred codes are in the dataset.

| # | Journal | Publisher | Year | Owner | Soc. | Primary trigger | Ed. repl. | Successor | Rev. |
|---|---|---|---|---|---|---|---|---|---|
| 1 | Medical Journal of Australia | Australian Medical Association | 2015 | Society | 1 | Editor replacement | 1 | none | 0 |
| 2 | Lingua | Elsevier | 2015 | Publisher | 0 | Fees / OA | 0 | launched: Glossa | 0 |
| 3 | Scientific Reports | Springer Nature | 2017 | Publisher | 0 | Other | 0 | none | 1 |
| 4 | Journal of Algebraic Combinatorics | Springer Nature | 2017 | Publisher | 0 | Fees / OA | 0 | launched: Algebraic Combinatorics | 0 |
| 5 | Third World Quarterly | Taylor & Francis | 2017 | Independent | 0 | Editorial interference | 0 | none | 1 |
| 6 | International Journal of Occupational and Environmental Health | Taylor & Francis | 2017 | Publisher | 0 | Editorial interference | 1 | none | 0 |
| 7 | Building Research and Information | Taylor & Francis | 2018 | Publisher | 0 | Editor replacement | 1 | launched: Buildings & Cities | 0 |
| 8 | Diversity & Distributions | Wiley | 2018 | Publisher | 0 | Fees / OA | 0 | none | 0 |
| 9 | Nutrients | MDPI | 2018 | Publisher | 0 | Volume pressure | 0 | none | 0 |
| 10 | Journal of Molecular Medicine | Springer Nature | 2018 | Publisher | 0 | Editor replacement | 1 | none | 0 |
| 11 | Journal of Informetrics | Elsevier | 2019 | Publisher | 0 | Fees / OA | 0 | launched: Quantitative Science Studies | 0 |
| 12 | Revue d'intelligence artificielle | IIETA | 2019 | Publisher | 0 | Editorial interference | 0 | launched: Revue Ouverte d'Intelligence Artificielle (ROIA) | 0 |
| 13 | Ingénierie des Systèmes d'Information | IIETA | 2019 | Publisher | 0 | Editorial interference | 0 | launched: Revue Ouverte d'Ingénierie des Systèmes d'Information (ROISI) | 0 |
| 14 | European Law Journal | Wiley | 2020 | Publisher | 0 | Contract terms | 0 | launched: European Law Open | 0 |
| 15 | Acta Médica Peruana | Colegio Médico del Perú | 2021 | Society | 1 | Editorial interference | 1 | none | 0 |
| 16 | Molecular Genetics & Genomic Medicine | Wiley | 2021 | Publisher | 0 | Other | 0 | none | 1 |
| 17 | Aging Cell | Wiley / Anat. Soc. | 2022 | Society | 1 | Contract terms | 0 | none | 0 |
| 18 | Indoor Air | Wiley / ISIAQ | 2022 | Publisher | 1 | Fees / OA | 0 | launched: Indoor Environments | 0 |
| 19 | Journal of Biogeography | Wiley | 2023 | Publisher | 0 | Fees / OA | 1 | none | unclear |
| 20 | Kurdish Studies | Transnational Press | 2023 | Publisher | 1 | Ownership change | unclear | launched: Kurdish Studies Journal | 0 |
| 21 | Critical Public Health | Taylor & Francis | 2023 | Publisher | 0 | Fees / OA | 0 | launched: Journal of Critical Public Health | 0 |
| 22 | NeuroImage and NeuroImage: Reports | Elsevier | 2023 | Publisher | 0 | Fees / OA | 0 | launched: Imaging Neuroscience | 0 |
| 23 | Design Studies | Elsevier | 2023 | Publisher | 1 | Editor replacement | 1 | planned | 0 |
| 24 | British Journal of Psychiatry | CUP / RCPsych | 2023 | Society | 1 | Editorial interference | 0 | none | 0 |
| 25 | Journal of Geometric Mechanics | AIMS | 2023 | Publisher | 0 | Fees / OA | 0 | launched: Geometric Mechanics | 0 |
| 26 | Publications | MDPI | 2023 | Publisher | 0 | Volume pressure | 0 | none | unclear |
| 27 | eLife | Independent | 2023 | Independent | 0 | Other | 1 | none | 0 |
| 28 | Journal of International Students | Independent | 2023 | Independent | 0 | Fees / OA | 0 | none | 0 |
| 29 | The Journal of Political Philosophy | Wiley | 2023 | Publisher | 0 | Editor replacement | 1 | launched: Political Philosophy | 0 |
| 30 | Sustainable Organisations section of Frontiers in Sustainability | Frontiers | 2023 | Publisher | 0 | Volume pressure | 0 | none | unclear |
| 31 | Econometrics | MDPI | 2023 | Publisher | 0 | Other | 0 | none | 0 |

| # | Journal | Publisher | Year | Owner | Soc. | Primary trigger | Ed. repl. | Successor | Rev. |
|---|---|---|---|---|---|---|---|---|---|
| 32 | Aquatic Conservation: Marine and Freshwater Ecosystems | Wiley | 2023 | Publisher | 0 | Editor replacement | 1 | none | 0 |
| 33 | Theory and Society | Springer Nature | 2024 | Publisher | 0 | Editor replacement | 1 | launched: Theory and Social Inquiry | 0 |
| 34 | Journal of Economic Surveys | Wiley | 2024 | Publisher | 0 | Volume pressure | 0 | none | 0 |
| 35 | BMC Pregnancy and Childbirth | Springer Nature | 2024 | Publisher | 0 | Other | 0 | none | unclear |
| 36 | Gender, Work and Organization | Wiley | 2024 | Publisher | 0 | Editor replacement | 1 | unclear: Feminism and Organization | 0 |
| 37 | Syntax | Wiley | 2024 | Publisher | 0 | Other | 0 | launched: Syntactic Theory and Research | 0 |
| 38 | Philosophy & Public Affairs | Wiley | 2024 | Publisher | 0 | Fees / OA | 0 | launched: Free & Equal: A Journal of Ethics and Public Affairs | 0 |
| 39 | Intelligence | Elsevier | 2024 | Publisher | 1 | Editor replacement | 1 | launched: Intelligence & Cognitive Abilities | 0 |
| 40 | Journal of Human Evolution | Elsevier | 2024 | Publisher | 0 | Other | 0 | none | 0 |
| 41 | Psychotherapy and Psychosomatics | Karger | 2024 | Publisher | 1 | Editor replacement | 1 | none | 0 |
| 42 | Journal of Clinical Immunology | Springer Nature | 2024 | Publisher | 0 | Volume pressure | 0 | launched: Journal of Human Immunity | 0 |
| 43 | Sedimentology | Wiley | 2025 | Society | 1 | Contract terms | 0 | none | 0 |
| 44 | Journal of Architectural Education | Taylor & Francis | 2025 | Society | 1 | Editorial interference | 1 | none | 0 |
| 45 | Mathematical Logic Quarterly | Wiley | 2025 | Publisher | 1 | Editorial interference | 0 | launched: ZML: Zeitschrift für Mathematische Logik und Grundlagen der Mathematik | 0 |
| 46 | Amsterdam University Press book series | AUP / T&F | 2025 | Publisher | 0 | Ownership change | 0 | none | 0 |
| 47 | Criminology | Wiley | 2025 | Society | 1 | Other | 0 | none | 0 |
| 48 | GigaScience | Oxford University Press | 2025 | Publisher | 0 | Editor replacement | 1 | none | 0 |
| 49 | Journal of Philosophical Logic | Springer Nature | 2025 | Publisher | 0 | Fees / OA | 0 | launched: Philosophical Logic | 0 |
| 50 | Revue des Composites et des Matériaux Avancés | IIETA | 2026 | Publisher | 0 | Volume pressure | 0 | none | 0 |
| 51 | Communications in Algebra | Taylor & Francis | 2026 | Publisher | 0 | Editorial interference | 1 | none | 0 |
| 52 | Journal of Approximation Theory | Elsevier | 2026 | Publisher | 0 | Editor replacement | 1 | none | 0 |
| 53 | Natural Language Semantics | Springer Nature | 2026 | Publisher | 0 | Volume pressure | 0 | launched: Semantics of Natural Languages | 0 |
| 54 | Communications on Pure and Applied Mathematics | Wiley | 2026 | Unclear | 1 | Editorial interference | 0 | launched: Courant Journal of Pure and Applied Mathematics | 0 |
| 55 | Statistics and Computing | Springer Nature | 2026 | Publisher | 0 | Fees / OA | 0 | none | 0 |
| 56 | Review of Social Economy | Taylor & Francis | 2026 | Society | 1 | Fees / OA | 0 | planned: Journal of Social Economy and Philosophy | 0 |
| 57 | Games and Economic Behavior | Elsevier | 2026 | Publisher | 1 | Editor replacement | 1 | planned | 0 |

## Appendix B. Editor tenure dataset

The tenure dataset records 549 spells in the top editorial position at 110 journals (65 Elsevier, 45 other), active at any time from 2005 to 2020, collected in two rounds; the second round targeted Elsevier journals with gaps. Each row gives the journal, publisher in 2010, owner type, society affiliation, editorial structure, editor, title, start and end years, end reason where stated, whether the editor founded the journal, sources, and notes on uncertainty. Rows that could not be verified from a public source are flagged and excluded from the tests. Coverage is uneven: it is good for journals with published editor histories, such as society journals and journals whose editors publish their curricula vitae, and weak for journals whose editorial boards are documented only on publisher websites. The dataset will be posted with the paper.

## Appendix C. The model

This appendix sets out the model summarized in Section 2. The baseline in Appendices C.1 to C.5 is deliberately minimal. Appendices C.6 and C.7 extend it to collective action among board members, to public protest as a commitment, and to reversal after resignations.

### *C.1 Setup*

An owner, O, holds the title of a journal whose reputation is worth $R$. The reputation is embodied in the editors and board, B, who act collectively. The owner considers replacing the editor in order to implement a policy, such as higher publication volume or a new review model, that yields the owner a private gain $G > 0$ and imposes a cost $C \geq 0$ on the board, for example by lowering the standing of the work that board members certify. If the board exits, it launches a successor journal at cost $F \geq 0$ and takes with it a share $\theta \in [0, 1]$ of the reputation. The board values reputation under its own control at $w$ per unit, and the owner loses $m$ per unit of reputation that leaves, so the owner's loss from exit is $L = \theta m R$. The cost $F$ captures everything that makes starting again hard: finding a publisher or host, financing, indexing, and coordinating the board. A society that adopts the successor, or a non-profit platform that hosts it, lowers $F$; a society that designates the journal as its own, and moves with the board, raises $\theta$.

The timing is as follows (Figure 1 in Section 2). (1) The owner announces a replacement. (2) The board may protest. (3) The owner proceeds or withdraws. (4) If the owner proceeds, the board stays or exits; exit is irreversible once the successor is launched. With the status quo normalized to zero, the payoffs to the owner and the board are (0, 0) if the owner withdraws; $(G, -C)$ if it proceeds and the board stays; and $(G - L, \theta w R - F)$ if it proceeds and the board exits.

### *C.2 Complete information*

At stage 4 the board exits if and only if its payoff from exit exceeds its payoff from staying:

$$F < F* \equiv \theta wR + C.$$

**Proposition 1.** With complete information: (a) if $F \geq F*$, the owner proceeds and the board stays; (b) if $F < F*$ and $G < L$, the owner withdraws; and (c) if $F < F*$ and $G \geq L$, the owner proceeds and the board exits. In no case does the owner reverse a replacement after the board has exited.

*Proof.* By backward induction, the board exits at stage 4 if and only if $F < F*$. At stage 3 the owner compares 0 with $G$ if the board will stay and with $G - L$ if it will exit, which gives (a) to (c). Because exit is irreversible, a reversal after exit cannot recover the reputation, and the owner, having chosen to proceed knowing the board would leave, gains nothing from reversing. ■

Case (a) is a replacement that goes ahead because the board has nowhere to go; individual members may resign, but there is no collective exit and no successor. Case (b) is voice succeeding in Hirschman's (1970) sense: the threat of exit is credible and costly enough that the owner withdraws before anyone resigns. Case (c) is exit: the owner values its policy more than the reputation it loses.

### *C.3 Incomplete information*

In practice the owner does not observe the board's cost of starting again, which depends on whether a society or a non-commercial host will back a successor and whether the board can coordinate. Suppose the board knows $F$ but the owner believes that $F$ is distributed with a continuous cumulative distribution function $H$. Protest at stage 2 is then cheap talk, since a board with a high $F$ would also claim that it will leave, and the owner proceeds if and only if its expected payoff is non-negative:

$$G - H(F*) \cdot L \geq 0, \text{ that is, } H(F*) \leq G / L.$$

**Proposition 2.** With incomplete information: (a) the set of journals at which the owner proceeds shrinks as the share of reputation the board can take ($\theta$), the board's valuation of it ($w$), or the board's cost of staying ($C$) rises, and expands as the owner's gain ($G$) rises; (b) conditional on the owner proceeding, the board exits with probability $H(F*)$, and such exits can occur even when $G < L$, in which case the owner regrets the replacement ex post but, in the baseline in which exit is immediate, cannot reverse it; and (c) a fall in the cost of launching successors, in the sense of a first-order stochastically dominated shift in the distribution of $F$, reduces the set of journals at which the owner proceeds and raises the probability of exit at those where it does.

*Proof.* $F*$ is increasing in $\theta$, $w$, and $C$, and $H$ is non-decreasing, so $H(F*)$ is non-decreasing in each; $L = \theta mR$ is also increasing in $\theta$. The owner's condition $H(F*) \leq G/L$ is therefore tightened by higher θ, $w$, or $C$ and relaxed by higher $G$, which gives (a). Part (b) follows from the board's exit rule at stage 4 and the irreversibility of exit. For (c), a first-order stochastically dominated shift lowers the distribution of $F$ and so raises $H(F*)$ at every value of $F*$, which tightens the owner's condition and raises the exit probability. ■

**Lemma 1.** Suppose that $F$ is drawn independently of (θ, $w$, $C$, $G$, $R$). Then, among journals at which the owner proceeds, the probability that the board exits is $H(F*)$, which is non-decreasing in θ, $w$, and $C$, whatever the joint distribution of (θ, $w$, $C$, $G$, $R$) across journals.

*Proof.* The owner's decision depends on $F$ only through its distribution $H$, so conditioning on the decision to proceed does not change the distribution of $F$ at a journal with given parameters. The probability of exit there is $\Pr(F < F*) = H(F*)$, which is non-decreasing in θ, $w$, and $C$ because $F*$ is increasing in each. ■

Lemma 1 matters for the tests in Section 5, which observe only episodes in which the owner proceeded. Selection on the owner's decision changes which journals are observed, since a high θ deters replacements (Proposition 2(a)), but it does not change the probability of exit at a journal with a given θ. Comparisons of exit rates across observed episodes that differ in θ or in the cost of starting again therefore recover the sign of the effects in Proposition 2(b) and (c). The lemma fails if the owner knows more about $F$ than $H$ conveys, for example through private contact with a society, in which case observed replacements are tilted toward boards the owner expected to stay, and the observed differences understate the true ones.

### *C.4 Rules enforced at the owner's discretion*

Suppose the owner announces a general rule, such as a cap on the tenure of chief editors, but retains discretion over its enforcement. Enforcing the rule at a journal is equivalent to proceeding with a replacement there, so the owner enforces it only where $H(F*) \leq G/L$.

**Corollary.** A rule enforced at the owner's discretion is applied where boards' threats to exit are weak, with low θ, $w$, or $C$, or where the owner's gain from a change of policy is high, and is waived where boards are strong.

### *C.5 Editor tenure and rules on tenure*

The Corollary can be sharpened by recognizing that part of a journal's reputation attaches to its editor, and that this part grows with the editor's tenure. Let the share of reputation that leaves if the editor and board exit be $\theta(\tau)$, increasing in the editor's tenure $\tau$, so that $F*(\tau) = \theta(\tau)wR + C$ and $L(\tau) = \theta(\tau)mR$ are both increasing in $\tau$. Suppose the owner announces a cap $\bar{\tau}$ on the tenure of chief editors.

**Proposition 3.** (a) If the cap is binding, every editor who reaches $\bar{\tau}$ is replaced, so the probability of leaving at the cap is one, whatever the strength of the board. (b) If the cap is enforced at the owner's discretion, an editor who reaches $\bar{\tau}$ is replaced if and only if $H(F*(\bar{\tau})) \cdot L(\bar{\tau}) \leq G$. The probability of replacement at the cap is then below one, lower where the board can carry more of the reputation, and higher where the owner's gain from a change of policy is larger. (c) If the renewal decision recurs at the end of each contract after $\bar{\tau}$, then under discretion the editors least likely to be replaced are the longest-serving, the opposite of what a binding cap implies.

*Proof.* Part (a) is immediate. Part (b) applies the owner's condition from Appendix C.3 at $\tau = \bar{\tau}$ across journals that differ in $\theta$, $w$, $C$, and $G$. For (c), at each renewal after $\bar{\tau}$ the owner applies the same condition at the current $\tau$; $H(F*(\tau)) \cdot L(\tau)$ is increasing in $\tau$ because $\theta(\tau)$ is, so, holding $G$ fixed, the condition is harder to satisfy the longer the editor has served. If instead the owner's gain from a change rises with tenure, as the stated rationale that long tenures settle into the status quo suggests, (c) need not hold. ■

Proposition 3 gives P3 a form that can be tested. Under a binding cap, essentially all chief editors with ten or more years of tenure while the cap was in force should have left by the end of that period. Under discretionary enforcement, few should have left, and those who did should be concentrated at journals with weak boards.

### *C.6 Collective exit as a coordination problem*

The baseline treats the board as a single actor. Suppose instead that it has many members, none of them pivotal. A member who leaves receives $v - e(\text{i})$ if the successor is viable and $-e(\text{i})$ if it is not, where $v > 0$ is the value to her of a place on a viable successor and $e(\text{i}) \geq 0$ is her private cost of moving. A member who stays receives $-c$, the cost to her of the owner's policy. The successor is viable if at least a share $k$ of members leave. Let $S(x)$ denote the share of members with $e(\text{i}) < x$.

**Proposition 4.** (a) If $S(c) < k \leq S(v + c)$, there are two equilibria. In one, the members with $e(\text{i}) < v + c$ leave and found a successor. In the other, only the members with $e(\text{i}) < c$ leave, one by one, and no successor forms. (b) If a society or host commits in advance to support the successor whatever the number who leave, leaving is a dominant strategy for every member with $e(\mathrm{i}) < v + c$, and the first equilibrium is the unique outcome.

*Proof.* (a) If the successor will be viable, a member leaves if and only if $v - e(\mathrm{i}) > -c$, that is, $e(\mathrm{i}) < v + c$; the share leaving is $S(v + c) \geq k$, so the successor is indeed viable. If it will not be viable, a member leaves if and only if $-e(\mathrm{i}) > -c$, that is, $e(\mathrm{i}) < c$; the share leaving is $S(c) < k$, so the successor is indeed not viable. (b) With viability guaranteed, each member compares $v - e(\mathrm{i})$ with $-c$ whatever the others do. ■

Proposition 4 is a second channel, alongside θ, through which societies and non-commercial hosts raise the board's outside option: by committing in advance, they remove the coordination failure. It implies that board departures should be bimodal: near-total exits with successors where an organization commits to the successor, and scattered resignations without successors where none does. A public letter signed by most of the board, of the kind that accompanies many mass resignations, serves a similar purpose by making each member's intention known to the others. GEB, where the editors and 58 of 62 advisory editors left together and the society later ended its association with the title, and FRL, where board members resigned individually and no successor formed, illustrate the two equilibria.

### *C.7 Voice as commitment, and reversal after resignations*

In the baseline, protest is cheap talk and plays no role in equilibrium. Protest matters if it commits the board. Suppose that before the owner decides, the members can make a public pledge to resign if the owner proceeds, and that breaking the pledge costs the board $d > 0$ in credibility. A pledge changes the board's own exit rule: after pledging, the board exits if and only if $\theta wR - F \geq -C - d$, that is, if and only if $F < F* + d$.

**Proposition 5.** With complete information, a board with $F* \leq F < F* + d$ pledges, and the owner withdraws, if and only if $G < L$. A board with $F < F*$ needs no pledge, and a board with $F \geq F* + d$ cannot make exit credible. A public pledge therefore enlarges the set of boards whose voice succeeds from those with $F < F*$ to those with $F < F* + d$.

*Proof.* After a pledge, the board exits if and only if $F < F* + d$, so the owner withdraws if and only if $G < L$. For a board with $F* \leq F < F* + d$, pledging yields 0 when $G < L$, against $-C$ without a pledge.

When $G \geq L$ the owner proceeds anyway, and the board, having pledged, exits with $\theta wR - F \leq -C$, so it does not pledge. A board with $F \geq F* + d$ would stay after pledging and pay $d$, so its pledge is not credible and it does not make one. ■

With incomplete information, a pledge is informative because it is costly to a board that would not carry it out: such a board pays $d$ if the owner calls its bluff. The prediction is that successful voice is public and collective, takes the form of commitments to resign, and comes before any resignation.

The baseline also treats exit as irreversible the moment the board resigns. In practice there is a window between resignations and the launch of a successor; at GEB, the resigning editors gave three months' notice. Suppose that if the owner reverses within that window, by reinstating the editor or withdrawing the policy, it recovers a share $\rho \in [0, 1]$ of the reputation that would otherwise leave, gives up $G$, and incurs a cost $K \geq 0$, which includes commitments to incoming editors and the signal that a reversal sends to the boards of the owner's other journals.

**Proposition 6.** (a) An owner that has proceeded and faces resignations reverses if and only if $\rho L > G + K$. (b) With complete information, resignations occur only if $G \geq L$, so reversal never occurs. (c) With incomplete information, the owner proceeded only if $G \geq H(F*)L$, so reversal requires $\rho > H(F*) + K/L$: it occurs only where the owner underestimated the board's ability to leave, much of the reputation can still be recovered, and the owner's gain from its policy is small relative to the reputation at stake.

*Proof.* After resignations, the owner compares $G - L$ from holding firm with $-K - (1 - \rho)L$ from reversing, which gives (a). For (b), $G \geq L$ and $\rho \leq 1$ imply $\rho L \leq G + K$. For (c), combine (a) with the owner's condition $G \geq H(F*)L$. ■

Proposition 6 gives the timing prediction content beyond the baseline. Reversals after resignation are possible, but they should be confined to disputes in which the owner's gain is small, such as a dispute over a single paper or decision, and should be absent where the change is a new editor chosen to carry out a policy, where $G$ is large.